\pdfoutput=1
\documentclass[10pt,twocolumn,letterpaper]{article}
\usepackage[letterpaper,textwidth=516pt,columnsep=12pt,top=0.75in,bottom=0.9in]{geometry}
\usepackage[T1]{fontenc}

\usepackage{amsmath}
\usepackage{graphicx}
\usepackage{booktabs}
\usepackage{xcolor}
\usepackage{microtype}
\usepackage[numbers,sort&compress]{natbib}
\usepackage[hidelinks]{hyperref}
\newcommand{\Description}[1]{}

\renewcommand{\thesection}{\Roman{section}}
\renewcommand{\thesubsection}{\mbox{\thesection-\Alph{subsection}}}

\makeatletter
\renewcommand\subsection{\@startsection{subsection}{2}{\z@}%
  {-3.25ex\@plus -1ex \@minus -.2ex}{1.5ex \@plus .2ex}%
  {\normalfont\large\bfseries\raggedright}}
\makeatother

\newif\ifsubmission
\submissionfalse
\newif\ifsupplementary
\supplementaryfalse

\usepackage{array}
\newcolumntype{L}[1]{>{\raggedright\arraybackslash}p{#1}}
\usepackage{tikz}
\usepackage{pgfplots}\pgfplotsset{compat=1.18}
\usepgfplotslibrary{fillbetween}
\usetikzlibrary{arrows.meta,positioning,fit,calc}
\usepackage{cleveref}
\crefname{section}{Section}{Sections}
\Crefname{section}{Section}{Sections}

\microtypesetup{expansion=false}

\AtBeginDocument{\raggedbottom}

\newcommand{\Sys}[0]{}
\usepackage{xspace}
\xspaceaddexceptions{\%,\times}

\newcommand{\RhoAtt}{\rho_{\mathrm{att}}}
\newcommand{\RhoLay}{\rho_{\mathrm{lay}}}
\newcommand{\RhoFfn}{\rho_{\mathrm{ffn}}}
\newcommand{\RhoNorm}{\rho_{\mathrm{nrm}}}
\newcommand{\CostLawGeneral}{B(n) = \RhoAtt L C n + (\RhoLay + \RhoFfn K) L C - (C - W) + \tau}
\newcommand{\LawSlope}{48}
\newcommand{\LawColConst}{305}
\newcommand{\LawPromptConst}{34}
\newcommand{\LawGenConst}{36}
\newcommand{\PromptStepBoots}{3716\xspace}
\newcommand{\GenStepBoots}{3814\xspace}
\newcommand{\CachedStepBoots}{436\xspace}
\newcommand{\DeadWorkGenBoots}{3766\xspace}
\newcommand{\PackedGenBoots}{742\xspace}
\newcommand{\DeadWorkGenSecDerived}{1387\xspace}
\newcommand{\LlamaDeadGenBoots}{62{,}365\xspace}
\newcommand{\LlamaPackedGenBoots}{19{,}357\xspace}
\newcommand{\LlamaDeadGenSec}{22{,}975\xspace}
\newcommand{\LlamaPackedGenSec}{7{,}131\xspace}
\newcommand{\RefreshPriceMs}{368.4\xspace}
\newcommand{\StockSecPerToken}{1437.3\xspace}
\newcommand{\CachedSecPerGenStep}{187\xspace}
\newcommand{\StockSecPerGenStep}{1456\xspace}
\newcommand{\PackedSecPerGenStep}{345\xspace}
\newcommand{\GptTwoSecPerToken}{1086.1\xspace}
\newcommand{\GptTwoLawPromptBoots}{11458\xspace}
\newcommand{\GptTwoLawGenBoots}{11748\xspace}
\newcommand{\GptTwoMeasPromptBoots}{11461\xspace}
\newcommand{\GptTwoMeasGenBoots}{11751\xspace}
\newcommand{\LlamaPromptBoots}{90{,}779\xspace}
\newcommand{\LlamaGenBoots}{93{,}085\xspace}
\newcommand{\LlamaPromptHours}{9.29\xspace}
\newcommand{\LlamaGenHours}{9.53\xspace}
\newcommand{\SecurityBits}{128\xspace}
\newcommand{\SecurityAttackLogRop}{140.8}
\newcommand{\HorizonDivergent}{48\xspace}
\newcommand{\HorizonExactAgainTo}{52\xspace}
\newcommand{\CacheAgingExponent}{0.814\xspace}
\newcommand{\AlignmentFloor}{0.997\xspace}
\newcommand{\ALIGNMENTFLOOR}{\AlignmentFloor}
\newcommand{\SoftmaxScalar}{1.10\xspace}
\newcommand{\SoftmaxPostScalarFloorRatio}{0.63\xspace}
\newcommand{\DevGridStepOne}{337\xspace}
\newcommand{\DevStepCutLowPct}{11\xspace}
\newcommand{\DevStepCutHighPct}{13\xspace}
\newcommand{\DevThreeCutGridPct}{11\xspace}
\newcommand{\DevThreeCutRepPct}{9\xspace}
\newcommand{\DevRepBarrierWallTwo}{41410.1\xspace}
\newcommand{\DevRepBarrierWallThree}{43670.3\xspace}
\newcommand{\DevRepBarrierWallFour}{38265.7\xspace}
\newcommand{\DevReplicateTwoPct}{2.0\xspace}
\newcommand{\DevReplicateFourPct}{0.2\xspace}
\newcommand{\DevThreeExcessPct}{8.8\xspace}
\newcommand{\DevRepWidthTwoMeanTwo}{363.5\xspace}
\newcommand{\DevRepWidthTwoMeanThree}{426.3\xspace}
\newcommand{\DevRepWidthTwoMeanFour}{373.2\xspace}
\newcommand{\DevRepRatioTwo}{1.0822\xspace}
\newcommand{\DevRepRatioThree}{1.1412\xspace}
\newcommand{\DevRepSecPerTokenOne}{255.1\xspace}
\newcommand{\DevRepSecPerTokenTwo}{234.5\xspace}
\newcommand{\DevRepSecPerTokenThree}{242.4\xspace}
\newcommand{\DevRepSecPerTokenFour}{235.4\xspace}
\newcommand{\DevStepWallSpreadPct}{3.7\xspace}
\newcommand{\DevWidthTwoEarlyCount}{51\xspace}
\newcommand{\DevWidthTwoEarlyLowMs}{376\xspace}
\newcommand{\DevWidthTwoEarlyHighMs}{430\xspace}
\newcommand{\DevWidthTwoLateCount}{26\xspace}
\newcommand{\DevWidthTwoLateLowMs}{449\xspace}
\newcommand{\DevWidthTwoLateHighMs}{514\xspace}
\newcommand{\DevGridWidthTwoLateLowMs}{381\xspace}
\newcommand{\DevGridWidthTwoLateHighMs}{406\xspace}
\newcommand{\DevRepThreeStepExcessS}{11\xspace}
\newcommand{\DevRepThreeOutsideBarriersS}{8.7\xspace}
\newcommand{\DevPromptStepThreeS}{178\xspace}
\newcommand{\DevPromptStepLowS}{174\xspace}
\newcommand{\DevPromptStepHighS}{175\xspace}
\newcommand{\CeilMasterS}{258\xspace}
\newcommand{\CeilStepFloorS}{291\xspace}
\newcommand{\CeilFactor}{1.16\xspace}
\newcommand{\CeilAboveFloorLowPct}{0.5\xspace}
\newcommand{\CeilAboveFloorHighPct}{5.3\xspace}
\newcommand{\DevModelRunOnePct}{1.5\xspace}
\newcommand{\DevModelRunTwoPct}{3.4\xspace}
\newcommand{\MatmulMultiDevErr}{1.0e+01\xspace}
\newcommand{\MatmulTol}{5e-02\xspace}
\newcommand{\MatmulSingleDevErr}{1.3e-07\xspace}
\newcommand{\GptTwoStockSecPerToken}{4668.5\xspace}
\newcommand{\GptTwoStockSecPerGenStep}{4715\xspace}
\newcommand{\GptTwoDeadSecPerGenStep}{4059\xspace}
\newcommand{\GptTwoPackedSecPerGenStep}{1331\xspace}
\newcommand{\GptTwoCachedSecPerGenStep}{544\xspace}
\newcommand{\GptTwoPromptStepSec}{4630\xspace}
\newcommand{\GptTwoTokensPerHour}{6.6\xspace}
\newcommand{\GptTwoPromptStepHours}{1.29\xspace}
\newcommand{\GptTwoDeadCountPct}{30.4\xspace}
\newcommand{\GptTwoDeadWallPct}{29.7\xspace}
\newcommand{\GptTwoPackCountPct}{81.7\xspace}
\newcommand{\GptTwoPackWallPct}{76.1\xspace}
\newcommand{\GptTwoCachedStepBoots}{1072\xspace}
\newcommand{\GptTwoCachedStepBootsFour}{1168\xspace}
\newcommand{\GptTwoLawReadings}{20\xspace}
\newcommand{\GptTwoLawContexts}{10\xspace}
\newcommand{\GptTwoTokenCostTwoCol}{192\xspace}
\newcommand{\GptTwoTokenCostThreeCol}{288\xspace}
\newcommand{\GptTwoDeadGenBoots}{10023\xspace}
\newcommand{\GptTwoPackedPromptReadings}{ten\xspace}
\newcommand{\GptTwoPackedGenReadings}{ten\xspace}
\newcommand{\GptTwoPackedGenHits}{five\xspace}
\newcommand{\GptTwoPackedGenMissLow}{192\xspace}
\newcommand{\GptTwoPackedGenMissHigh}{2880\xspace}
\newcommand{\GptTwoPackedGenBoots}{2535\xspace}
\newcommand{\GptTwoLayers}{12\xspace}
\newcommand{\GptTwoChunks}{4\xspace}
\newcommand{\GptTwoLogitCts}{4\xspace}
\newcommand{\GptTwoLawColConst}{731\xspace}
\newcommand{\GptTwoLawPromptConst}{52\xspace}
\newcommand{\GptTwoLawGenConst}{54\xspace}
\newcommand{\GptTwoLawSlopeCol}{96\xspace}
\newcommand{\GptTwoPackedMissWidth}{eight\xspace}
\newcommand{\TtftDeadSharePctLow}{45.8\xspace}
\newcommand{\TtftDeadSharePctHigh}{46.5\xspace}
\newcommand{\TtftPackSharePctLow}{80.6\xspace}
\newcommand{\TtftPackSharePctHigh}{83.9\xspace}
\newcommand{\TtftPriceGapPctLow}{3.0\xspace}
\newcommand{\TtftPriceGapPctHigh}{28.4\xspace}
\newcommand{\LlamaCachedBoots}{4{,}260\xspace}
\newcommand{\LlamaCachedHours}{0.44\xspace}
\newcommand{\LlamaCacheFactor}{21.9\xspace}
\newcommand{\ProjSlateShapeCount}{seven\xspace}
\newcommand{\ProjContextMax}{2048\xspace}
\newcommand{\ProjShortfallPctLow}{3.5\xspace}
\newcommand{\ProjShortfallPctHigh}{27.4\xspace}
\newcommand{\ProjTransferMs}{26.3\xspace}
\newcommand{\ProjTransferRateGiBs}{4.67\xspace}
\newcommand{\ProjKeySetGiB}{49.63\xspace}
\newcommand{\ProjMemErrPct}{five\xspace}
\newcommand{\ProjParkPairMiB}{12.0\xspace}
\newcommand{\ProjLlamaEightGenSec}{34{,}293\xspace}
\newcommand{\ProjLlamaEightCachedSec}{1{,}569\xspace}
\newcommand{\ProjLlamaSeventyGrowth}{42\xspace}
\newcommand{\ProjMistralGrowth}{31\xspace}
\newcommand{\HorizonReplicatedLengths}{2\xspace}
\newcommand{\CmpCalls}{30\xspace}
\newcommand{\CmpGeluMs}{70.35\xspace}
\newcommand{\CmpLnMs}{43.74\xspace}
\newcommand{\CmpMatmulSecPerCol}{0.2993\xspace}
\newcommand{\CmpMatmulReps}{three\xspace}
\newcommand{\CmpNexusMatmulSecPerCol}{1.31\xspace}
\newcommand{\CmpSoftmaxMs}{20.01\xspace}
\newcommand{\CmpGeluNexusMs}{69\xspace}
\newcommand{\CmpLnNexusMs}{45\xspace}
\newcommand{\CmpSoftmaxNexusMs}{20\xspace}
\newcommand{\InteractiveSavingPctRounded}{2.2\xspace}
\newcommand{\InteractiveClientHours}{4.72\xspace}
\newcommand{\GptTwoTwoStepFactor}{4.2\xspace}
\newcommand{\LlamaInteractiveRemoved}{64\xspace}
\newcommand{\LlamaInteractiveSavingPct}{1.5\xspace}
\newcommand{\LlamaInteractiveClientHours}{6.9\xspace}
\newcommand{\MistralGenBoots}{93{,}060\xspace}
\newcommand{\MistralDeadGenBoots}{62{,}340\xspace}
\newcommand{\MistralPackedGenBoots}{19{,}332\xspace}
\newcommand{\MistralCachedBoots}{4{,}235\xspace}
\newcommand{\MistralGenSec}{34{,}283\xspace}
\newcommand{\MistralDeadGenSec}{22{,}966\xspace}
\newcommand{\MistralPackedGenSec}{7{,}122\xspace}
\newcommand{\MistralCachedSec}{1{,}560\xspace}
\newcommand{\QwenEightGenBoots}{101{,}483\xspace}
\newcommand{\QwenEightDeadGenBoots}{66{,}923\xspace}
\newcommand{\QwenEightPackedGenBoots}{18{,}539\xspace}
\newcommand{\QwenEightCachedBoots}{4{,}434\xspace}
\newcommand{\QwenEightGenSec}{37{,}386\xspace}
\newcommand{\QwenEightDeadGenSec}{24{,}654\xspace}
\newcommand{\QwenEightPackedGenSec}{6{,}830\xspace}
\newcommand{\QwenEightCachedSec}{1{,}634\xspace}
\newcommand{\LlamaSeventyGenBoots}{439{,}333\xspace}
\newcommand{\LlamaSeventyDeadGenBoots}{275{,}493\xspace}
\newcommand{\LlamaSeventyPackedGenBoots}{91{,}173\xspace}
\newcommand{\LlamaSeventyCachedBoots}{15{,}668\xspace}
\newcommand{\LlamaSeventyGenSec}{161{,}850\xspace}
\newcommand{\LlamaSeventyDeadGenSec}{101{,}492\xspace}
\newcommand{\LlamaSeventyPackedGenSec}{33{,}588\xspace}
\newcommand{\LlamaSeventyCachedSec}{5{,}772\xspace}
\newcommand{\StockPromptStepSec}{1421\xspace}
\newcommand{\GptTwoStockPromptStepSec}{4622\xspace}
\newcommand{\LlamaEightLayers}{32\xspace}
\newcommand{\LlamaEightChunks}{4\xspace}
\newcommand{\LlamaEightLogitCts}{8\xspace}
\newcommand{\LlamaEightPositions}{4\xspace}
\newcommand{\LlamaEightLawSlope}{256\xspace}
\newcommand{\LlamaEightLawColConst}{1{,}887\xspace}
\newcommand{\LlamaEightLawPromptConst}{68\xspace}
\newcommand{\LlamaEightLawGenConst}{70\xspace}
\newcommand{\MistralLayers}{32\xspace}
\newcommand{\MistralChunks}{4\xspace}
\newcommand{\MistralLogitCts}{3\xspace}
\newcommand{\MistralPositions}{4\xspace}
\newcommand{\MistralLawSlope}{256\xspace}
\newcommand{\MistralLawColConst}{1{,}887\xspace}
\newcommand{\MistralLawPromptConst}{43\xspace}
\newcommand{\MistralLawGenConst}{45\xspace}
\newcommand{\QwenEightLayers}{36\xspace}
\newcommand{\QwenEightChunks}{3\xspace}
\newcommand{\QwenEightLogitCts}{10\xspace}
\newcommand{\QwenEightPositions}{4\xspace}
\newcommand{\QwenEightLawSlope}{288\xspace}
\newcommand{\QwenEightLawColConst}{1{,}763\xspace}
\newcommand{\QwenEightLawPromptConst}{78\xspace}
\newcommand{\QwenEightLawGenConst}{80\xspace}
\newcommand{\LlamaSeventyLayers}{80\xspace}
\newcommand{\LlamaSeventyChunks}{4\xspace}
\newcommand{\LlamaSeventyLogitCts}{8\xspace}
\newcommand{\LlamaSeventyPositions}{2\xspace}
\newcommand{\LlamaSeventyLawSlope}{640\xspace}
\newcommand{\LlamaSeventyLawColConst}{4{,}719\xspace}
\newcommand{\LlamaSeventyLawPromptConst}{68\xspace}
\newcommand{\LlamaSeventyLawGenConst}{70\xspace}
\newcommand{\QwenSeventyTwoLayers}{80\xspace}
\newcommand{\QwenSeventyTwoChunks}{4\xspace}
\newcommand{\QwenSeventyTwoLogitCts}{10\xspace}
\newcommand{\QwenSeventyTwoPositions}{2\xspace}
\newcommand{\QwenSeventyTwoLawSlope}{640\xspace}
\newcommand{\QwenSeventyTwoLawColConst}{4{,}719\xspace}
\newcommand{\QwenSeventyTwoLawPromptConst}{78\xspace}
\newcommand{\QwenSeventyTwoLawGenConst}{80\xspace}
\newcommand{\QwenSeventyTwoGenBoots}{439{,}343\xspace}
\newcommand{\QwenSeventyTwoDeadGenBoots}{275{,}503\xspace}
\newcommand{\QwenSeventyTwoPackedGenBoots}{91{,}183\xspace}
\newcommand{\QwenSeventyTwoCachedBoots}{15{,}678\xspace}
\newcommand{\QwenSeventyTwoGenSec}{161{,}854\xspace}
\newcommand{\QwenSeventyTwoDeadGenSec}{101{,}495\xspace}
\newcommand{\QwenSeventyTwoPackedGenSec}{33{,}592\xspace}
\newcommand{\QwenSeventyTwoCachedSec}{5{,}776\xspace}
\newcommand{\ProjFitsNanoCtx}{1024\xspace}
\newcommand{\ProjFitsGptTwoCtx}{512\xspace}
\newcommand{\ProjFitsEightBCtx}{64\xspace}

\newcommand{\suppappendix}[2]{%
\expandafter\gdef\csname suppapp@#1\endcsname{#2}}
\newcommand{\suppref}[1]{%
\ifcsname suppapp@#1\endcsname\csname suppapp@#1\endcsname\else ??\fi}

\ifsupplementary
\newcommand{\paperonly}[1]{}
\newcommand{\tronly}[1]{}
\newcommand{\supponly}[1]{#1}
\else
\ifsubmission
\newcommand{\paperonly}[1]{#1}
\newcommand{\tronly}[1]{}
\else
\newcommand{\paperonly}[1]{}
\newcommand{\tronly}[1]{#1}
\fi
\newcommand{\supponly}[1]{}
\fi

\renewcommand{\Sys}[0]{\mbox{AR-HE}}

\title{Static Bootstrap Placement for Encrypted Language Model Decoding}
\author{%
Halil \.{I}brahim Kanpak, Didem Unat%
\thanks{Halil \.{I}brahim Kanpak is with the Department of Computer
Engineering, Ko\c{c} University, \.{I}stanbul, T\"{u}rkiye (e-mail:
hkanpak21@ku.edu.tr).}%
\thanks{Didem Unat is with the Department of Computer Engineering, Ko\c{c}
University, \.{I}stanbul, T\"{u}rkiye (e-mail: dunat@ku.edu.tr).}}

\hypersetup{
  pdftitle={Static Bootstrap Placement for Encrypted Language Model Decoding},
  pdfauthor={Halil Ibrahim Kanpak, Didem Unat}}

\begin{document}

\maketitle

\begin{abstract}
Language models increasingly serve prompts that carry private data, and secure inference under homomorphic encryption lets a client outsource the computation without revealing the prompt. Existing secure inference systems run a forward pass without consuming a token under encryption, and generating text with them requires a client round trip at every generated token. Keeping the loop on the server instead requires selecting and consuming a token under encryption, and placing bootstraps for a loop body that grows with the context. We build \Sys{}, which runs the whole loop on the server, selects each token under encryption, retrieves its embedding, and writes it back into the encrypted state. The client sends one prompt and remains offline until the output.
One rule places every bootstrap in the run, without search, so the bootstrap cost of a token is a formula in the context length that is known before the run starts.
The schedule skips work whose result cannot reach the output, packs bootstraps that share an operand, and keeps the keys and values of past positions in an encrypted cache. With every optimization applied, generating a GPT-2 small token costs \GptTwoCachedSecPerGenStep seconds on one NVIDIA H100, down from \GptTwoStockSecPerGenStep seconds without optimization. The prompt step before it costs \GptTwoPromptStepSec seconds.  The cache alone takes a generated step from \GptTwoMeasGenBoots bootstraps to \GptTwoCachedStepBoots. The formula predicts every step we measured, including steps of a model it was not derived
from.

\end{abstract}

\section{Introduction}
\label{sec:intro}
Serving a language model is autoregressive generation, where each generated
token is appended to the context and becomes the input of the next step.
Under CKKS homomorphic encryption \cite{ckks} this loop is expensive. On one NVIDIA H100,
a generated token of GPT-2 small costs \GptTwoStockSecPerToken{} seconds
under encryption without any optimization. A recent tutorial~\cite{edbt} and
the authors of Hyperion~\cite{hyperion} list encrypted autoregressive
generation as an open research problem.

Prior work on private transformer inference has focused on the forward
pass, which is the body of the generation loop. It provides packing
layouts~\cite{moai, stip, ellmo}, matrix multiplication
kernels~\cite{jkls, thor, tricycle}, polynomial replacements for the
nonlinear stages~\cite{powerformer, leaf}, and GPU libraries and systems
fast enough to run whole models~\cite{phantom, encllm, sylph}. The three
closest systems, NEXUS, Cerium and Cachemir~\cite{nexus, cerium, cachemir},
do not feed a selected token back into the encrypted state. Cachemir
caches keys and values across decoding steps, but the tokens it decodes
are supplied from outside the encrypted computation instead of being
selected under encryption.
Generating text with these systems requires keeping the client actively in the loop. At every step, the client must decrypt the server's output logits, select the next token, and re-encrypt it back to the server. In our measurements, this client-assisted approach yields a negligible \InteractiveSavingPctRounded\% savings in server runtime per token, yet forces the client to stay online for \InteractiveClientHours hours just to generate 16 tokens of GPT-2 Small.
This defeats the primary premise of homomorphic encryption, which is
that the client offloads heavy computation to an untrusted server and goes offline.

Running the whole generation loop on the server is more than calling the forward pass repeatedly. Three problems arise, and each one comes from the loop. First, the loop body grows. Step $n$ attends over $n$ positions, so each step holds more ciphertexts and needs more bootstraps than the step before it. Because bootstraps dominate the cost of a token, prior work uses a placement tool to decide where a program bootstraps. DaCapo, Orion, ReSBM and Orbit~\cite{dacapo, orion, resbm, orbit} search over a program graph whose size is fixed before the search. HALO~\cite{halo} handles loops, but its body must use the same levels in every iteration. Neither fits a decode loop, whose body grows with every token. Second, the server must select the next token, retrieve its embedding and write it back into the state. Because the server may not learn which token it chose, all three operations run under encryption. Third, the cost of a run is not known before it starts. The usual approach bootstraps a ciphertext when it runs out of levels. The server then learns the level of a ciphertext only when it computes that ciphertext, and it cannot tell in advance which bootstraps a run needs or which of them could run in parallel.

We present \Sys{}, which runs the whole generation loop on the server. The client sends one encrypted prompt and receives encrypted tokens at the end of the run. \Sys{} is a layer between the generation loop and the CKKS GPU library Phantom \cite{phantom}, which performs the ring arithmetic, the key switching and the bootstrapping. At each step, the server selects the next token with CutMax's comparison-free argmax~\cite{cutmax}. The argmax yields an encrypted one-hot vector. The server multiplies this vector by the plaintext embedding table to retrieve the token's embedding, and writes the embedding into the block of the state reserved for the new token with a masked write. Because each of these operations processes every candidate token, the server completes the step without learning which token it selected.

With the client offline, \Sys{} computes the bootstrap schedule before the run starts. We place the bootstraps of the growing body with one rule and optimize the schedule that results. Because nothing is decrypted between steps, the level of every ciphertext follows from four public values, namely the model architecture, the packing layout, the context length and the step index. At the entry of each stage of a step, \Sys{} bootstraps an operand whose level is below the level that stage needs. Evaluating this rule over a step gives its bootstrap count as a closed form in the context length.

\Sys{} rewrites the schedule before the run and applies four optimizations. \emph{Skipping dead work} drops the bootstraps whose results cannot reach the output because the causal mask zeroes them or the context has not reached their state ciphertexts. \emph{Packing} sends operands with a few distinct values across thousands of slots through one ciphertext, so one bootstrap refreshes several operands. \emph{The device axis} runs independent bootstraps of the schedule on separate GPU devices. \emph{The cache} keeps earlier keys and values encrypted at the level attention reads, and every cached step we measured reads the cache without bootstrapping it. A generated step then costs a fixed number of bootstraps until the context fills a new state ciphertext.

We measure \Sys{} on NVIDIA H100 GPUs at \SecurityBits{} bits of classical security. On GPT-2 small with its full vocabulary, the four optimizations reduce a generated step from \GptTwoMeasGenBoots{} bootstraps to \GptTwoCachedStepBoots{}, and from \GptTwoStockSecPerGenStep{} seconds to \GptTwoCachedSecPerGenStep{}. The prompt step, which the cache does not shorten, costs \GptTwoPromptStepSec{} seconds. On nanoGPT, where the same three optimizations take a generated step to \CachedSecPerGenStep seconds, two and four devices reduce it by a further \DevStepCutLowPct{}\% to \DevStepCutHighPct{}\%. The formula matches the measured bootstrap count at every one of the \GptTwoLawReadings{} steps measured across \GptTwoLawContexts{} context lengths. We derived the formula from the circuit of nanoGPT, and it also predicts the counts of GPT-2 small. For Llama-3-8B at a context of 32, the formula gives \LlamaGenHours{} hours of bootstrap time for a generated step and \LlamaCachedHours{} hours with the cache.

This paper makes the following contributions.
\begin{itemize}
\item We present \Sys{}, the first system to run the whole autoregressive generation loop under CKKS on a GPU server. It selects the token, retrieves its embedding and writes it back under encryption, so the client stays offline between the prompt and the output.

\item We validate every encrypted decoding step against the plaintext model over the generated sequence. We report the steps at which the polynomial approximations of the nonlinear operators preserve the output token and the steps at which the argmax diverges.

\item We give a bootstrap placement rule for a loop whose body grows, and the closed-form cost model it yields, which predicts the bootstrap count of every step.

\item We build and evaluate four optimizations that the fixed schedule makes possible, skipping dead work, packing, an encrypted key and value cache, and running independent bootstraps on multiple GPUs.

\item We evaluate the full system on GPT-2 small and nanoGPT on one NVIDIA H100, and sweep the device axis up to four H100s on nanoGPT. We measure the operator
costs of Llama-3-8B, and project the cost of larger models such as Llama-3-70B.
\end{itemize}

Section~\ref{sec:background} states the threat model, introduces generation and CKKS, and measures the naive implementation of the loop. Section~\ref{sec:design} presents \Sys{}: the decode step, the rule that places the bootstraps, the closed form it yields, and the four optimizations. Section~\ref{sec:evaluation} describes the experimental setup and evaluates the system. Section~\ref{sec:related} discusses related work, and Section~\ref{sec:conclusion} concludes.
\section{Background and Motivation}

\label{sec:background}

In this section we give the threat model, the background on autoregressive generation and CKKS that the rest of the paper uses, and the cost of the loop without optimization.

\subsection{Secure inference and threat model}\label{sec:threat-model}

We consider a client that outsources language model inference to one server. The client holds the secret key. As in prior private inference frameworks under CKKS~\cite{nexus, cachemir}, the server is semi-honest and must not learn the prompt or any generated token. The client sends the prompt encrypted under CKKS together with the evaluation keys, and it can stay offline until it retrieves the encrypted output.
The server holds the model weights in plaintext, and it computes on ciphertexts, evaluation keys and public parameters only. We leave timing and memory access side channels out of scope, because the operation sequence and the memory addresses depend on public values only.

The server sees the public parameters, the evaluation keys, the encrypted prompt and every intermediate ciphertext of the run. Under the IND-CPA security of CKKS, this view is computationally indistinguishable from the same public parameters and keys together with encryptions of zero. The guarantee assumes circular security, that an encryption of the secret key under itself is safe to publish~\cite{gentry}, as every FHE deployment that bootstraps does. The model dimensions, the packing, the prompt length and the number of tokens to generate fix the operation sequence and the level of every ciphertext before the run, so the sequence of operations is independent of the encrypted values. The client decrypts only the final output and returns no decryption result to the server, so the key-recovery attack on approximate schemes~\cite{limicciancio}, which needs such results, does not apply. Appendix~\ref{sec:supp-security} gives the security reduction in full.

\subsection{Language models and autoregressive generation}
A decoder-only transformer generates text one token at a time. It processes the prompt once. Each later step runs the model over the context, which is the prompt and the tokens generated so far, and turns the final hidden state into logits over the vocabulary. It then selects the token with the largest logit and appends its embedding to the context. Step $n$ attends over all $n$ positions, so its work grows with the context. Plaintext serving systems keep the attention keys and values of past positions in a key and value cache, so each step computes them only for the new position. We call the step that processes the prompt the \textit{prompt step} and each later step a \textit{generated step}. Under encryption, the server sees the prompt, the context and every generated token only as ciphertexts.

\subsection{CKKS}
CKKS \cite{ckks} encrypts a vector of real numbers into one ciphertext over a polynomial ring of degree $N$, which holds up to $N/2$ slots that each carry one number. Addition and multiplication act slot by slot, and a rotation moves the slots by a fixed offset using a rotation key. Every multiplication consumes one level of the modulus chain, so a ciphertext at level $\ell$ supports $\ell$ more multiplications. The depth of a circuit is the number of levels it consumes. When a ciphertext runs out of levels, a bootstrap raises it to the top of the chain. A bootstrap itself consumes a fixed number of levels, so a parameter set leaves a fixed number of usable levels between bootstraps. A bootstrap also costs far more than any other operation. We use ring degree $2^{16}$ with $2^{14}$ slots in use, and a modulus chain that leaves 13 usable levels after a bootstrap, at \SecurityBits{} bits of classical security. One bootstrap costs \RefreshPriceMs{} ms on one H100. We build on Phantom \cite{phantom}, a GPU library for CKKS, and use NEXUS's bootstrap configuration \cite{nexus} unchanged.

\subsection{Motivation}
\label{sec:placement-problem}
\label{sec:generation-parts}

\begin{figure}[t]
\centering
\includegraphics{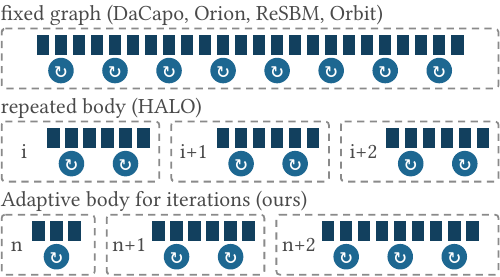}
\caption{Each tile is a ciphertext and each circled arrow a bootstrap. DaCapo, Orion, ReSBM and Orbit schedule a graph that is fixed before the search, and HALO schedules a body that repeats unchanged. The body of a decode step grows with the context.}
\Description{Three bands over one column. The top band draws a single
dashed container over one row of ciphertext tiles and the bootstraps under
them. The middle band draws three dashed containers of equal width holding
equal bodies, labeled i, i plus one and i plus two. The bottom band draws
three dashed containers of increasing width holding bodies of three, six
and nine tiles, labeled n, n plus one and n plus two.}
\label{fig:growingbody}
\end{figure}

To motivate \Sys{}, we first look at how a client generates text with the systems that exist today. Prior private transformer frameworks run language model inference one forward pass at a time, and the client takes part between passes. Each pass ends at the client, because these systems cannot feed a selected token back into the encrypted state, and the client therefore stays online for the whole generation. The server runs the model up to the logits and sends them to the client. The client decrypts them, selects the token, encrypts its embedding and returns it, and the server starts the next step. This avoids running the argmax under encryption. Generating 16 tokens on GPT-2 small keeps the client online for hours and saves the server a small share of its runtime.

To keep the client offline, the server itself needs to select each token and feed it back for the next generation step. A forward pass reads its input once. Generation feeds each output back as the next input, and under encryption that feedback is a ciphertext. Running this loop entirely under homomorphic encryption raises three problems.

The first problem is the growth of a step. Step $n$ attends over all $n$ earlier positions, so each step holds more ciphertexts and needs more bootstraps than the step before it. The second problem is the selection of the next token. Under encryption the server cannot learn which token it selected, so selection, retrieval and write-back each process every vocabulary entry and every state ciphertext. The third problem is the cost of a run. In a naive implementation on one NVIDIA H100 GPU, one generated token takes \StockSecPerToken{} seconds for nanoGPT and \GptTwoStockSecPerGenStep{} seconds for GPT-2 small. The bootstraps of a step set that time, since each costs three orders of magnitude more than a multiplication. The placement compilers take a program graph fixed before the search~\cite{dacapo, orion, resbm, orbit} or a loop body that consumes the same levels in every iteration~\cite{halo} (Figure~\ref{fig:growingbody}). Neither fits a decode loop, whose body grows with every generated token.

\section{System Design}
\label{sec:design}

\begin{figure*}[t] \centering \includegraphics[width=0.9\textwidth]{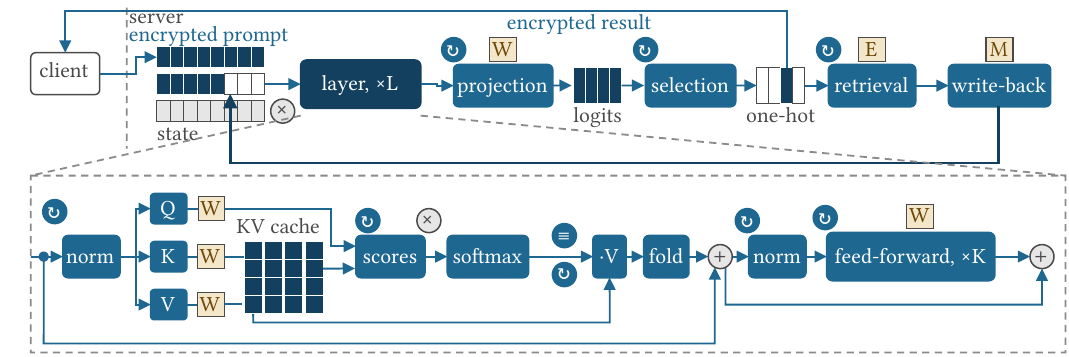} \caption{\Sys{} runs one decode step, and the panel below expands one layer. Tags mark what the server reads in plaintext: weights (W), the embedding table (E), the write-back mask (M). The circled arrow marks a bootstrap, the triple bar marks packing, and the cross marks skipped dead work.} \Description{A diagram of \Sys{} on one decode step, with the client left of a dashed rule and the server right of it. The top row carries the encrypted prompt through the layer stack, the projection, the selection, the retrieval and the write-back into the state. The panel below expands one layer into the norms, the query, key and value projections, the key and value cache, the scores, the softmax, the fold and the feed-forward.} \label{fig:sys} \end{figure*}

\Sys{} runs the transformer forward pass inside a decode loop and feeds the selected token back into the encrypted state, and in this section we present its design. Figure~\ref{fig:sys} shows one decode step, with its five stages above and one expanded layer below, and marks every bootstrap, the packed operands and the skipped dead work. The prompt arrives as a fixed set of ciphertexts that hold the residual stream of its positions, which we call the state, and the server writes each generated token into that set. We first describe the decode step, the work of producing one token, and its state layout, then the placement rule and the count it gives, then the four optimizations the fixed schedule makes possible.

\subsection{The decode step}\label{sec:decode-step}

A decode step takes the encrypted state of $n$ positions and produces one token in five stages (Figure~\ref{fig:sys}). The \textit{forward pass} runs the transformer layers over the state, and the \textit{projection} turns the final hidden vector into logits. \textit{Selection} produces an encrypted one-hot vector for the winning token, and \textit{retrieval} turns it into the token's embedding. \textit{Write-back} places the embedding into the state for the next step. The server passes the encrypted one-hot vector to retrieval and keeps a copy. The client receives the copies of all steps at the end of the run.

\paragraph{The state layout.} The encrypted state of a run is the residual stream of all its positions. A state ciphertext holds $b$ positions of it, one position per block of slots \cite{halevishoup}, so $b$ is the slot count divided by the width of a block, and the layout fixes the cost of a stage. A trivial layout allocates one state ciphertext per generated token. Each stage loops over the state ciphertexts, so the cost of a step would grow with the number of generated tokens. Instead, a run serving at most $n_{\max}$ positions reserves $C = \lceil n_{\max}/b \rceil$ state ciphertexts before the first step.  The reservation stays fixed for the run, so each step writes into a ciphertext that already exists.

Once every $b$ generated tokens, write-back starts the next reserved ciphertext, and at the other steps it writes into a ciphertext that already holds a generated token. Retrieval returns the embedding replicated across the blocks of one ciphertext. A plaintext mask keeps the copy in the block of the new position and zeroes the rest. Adding the result to the state ciphertext containing the block writes the token in place, so write-back has constant cost per step.

The server selects the token under encryption with CutMax's comparison-free argmax \cite{cutmax}. Retrieval multiplies the plaintext embedding table by the encrypted one-hot vector, and it processes all logit ciphertexts because the server must not learn which one holds the winner.

\subsection{The placement rule}\label{sec:placement-rule}

We place every bootstrap with one rule. Each stage has a threshold, the depth its operators consume. A site is a point where an operand enters a stage. At each site, the server bootstraps the operand if its level is below the threshold of the stage and passes it unchanged otherwise. The level at which an operand arrives depends only on public values. The whole schedule is therefore known before the run, and the bootstrap count of a step is a formula in the context length.

\begin{equation}\label{eq:cost-law}
\begin{split}
B(n) = {} & \rho_{att} L C n + (\rho_{lay} + \rho_{ffn} K) L C \\
& {} - (C - W) + (\tau + S + V_{ct} + w).
\end{split}
\end{equation}

Here $L$ is the layer count, $K$ the number of pieces the feed-forward layer is split into, $S$ the number of bootstraps required by selection, and $V_{ct}$ the number of logit ciphertexts. The tail $\tau$ is the once-per-step cost outside the layers for the final norm, logit product, selection, and write-back. $W$ is the number of state ciphertexts that already hold a generated token, and $w$ is one when this step writes into such a ciphertext. The coefficients $\rho_{att}$, $\rho_{lay}$ and $\rho_{ffn}$ count bootstrap sites in the attention path, in the part of a layer whose cost does not depend on the context, and in one piece of the feed-forward layer. We read each coefficient off the circuit.

\begin{figure}[t]
\centering
\includegraphics[width=\columnwidth]{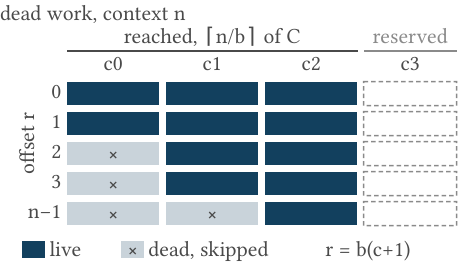}
\caption{Dead work on the state layout, at $b = 2$ positions per ciphertext and context $n = 5$. Attention scores each pair of a past position $r$ and a state ciphertext $c$, and each pair costs $\rho_{att}$ bootstraps per layer. The reservation stops at the $\lceil n/b \rceil$ state ciphertexts the context has reached, and causality zeroes each pair past $r = b(c+1)$, so the step skips those pairs.}
\Description{A grid of state ciphertexts against past positions, the reached ciphertexts bracketed, the reserved ciphertext dashed, the dead pairs crossed, and a stepped line marking the causal boundary.}
\label{fig:deadwork}
\end{figure}

\subsection{Dead work}\label{sec:dead-work}

Equation~\ref{eq:cost-law} counts all the bootstraps of a step, including those whose results never reach the output, and we call such work dead (Figure~\ref{fig:deadwork}). Attention computes a score for each pair of a position offset and a state ciphertext, and each pair has its own bootstraps. Two tests on public values determine whether a pair needs bootstraps. The reservation test stops the loop over the $C$ reserved ciphertexts at the $\lceil n/b \rceil$ the context has reached. The causality test skips pairs whose offsets lie past the last position in their ciphertext because the causal mask zeroes them. The server tests each pair before running its operations, so a dead pair is skipped before any of them run, and it bootstraps only the pairs on which the output depends.

The two tests act on different sites, so their savings add when both are enabled. The reservation test saves the bootstraps of one ciphertext until the context first reaches the last reserved ciphertext.

\paragraph{The cost of the mask.}  The mask costs one level on each pair, whether the output depends on it or not, and that level stays below all stage thresholds, so a cheaper mask would leave the bootstrap count unchanged.

\subsection{Packing}
 The second optimization packs the operands of several pairs into one ciphertext before a bootstrap, and one bootstrap then refreshes them all. The score of a pair is a sum over the dimensions of a head, and after that sum all slots of the head hold the same number. A ciphertext of attention weights therefore holds one distinct number per head in each block. A few dozen distinct values therefore occupy thousands of slots. The attention-value product requires replication, which we defer until after bootstrapping. We pack $G$ operands in one ciphertext, bootstrap the group once, and unpack after the bootstrap. Within the slots of one head, each operand occupies one slot and the rest hold zero. The packed ciphertext keeps the block layout of the state, so the masks of the unpacked circuit apply to it unchanged. Packing reuses the mask of the sum, and unpacking costs one level. Both come right after a bootstrap the rule has already placed, so the bootstrap count is unchanged. One plaintext holds the causal mask of all operands in a group, so the number of mask multiplications falls as the group grows. The bootstrap count of a step then becomes

\begin{equation}\label{eq:packed}
B_G(n) = \rho_{att} L \sum_{c} \left\lceil \frac{m_c}{G} \right\rceil + \gamma \, C_{live} + \tau',
\end{equation}

where $m_c$ is the number of past positions that state ciphertext $c$ attends over and $C_{live} = \lceil n/b \rceil$ is the number of state ciphertexts the context has reached. Here $\gamma$ and $\tau'$ are the per-ciphertext constant and the once-per-step constant of the formula, and the coefficients are those of Equation~\ref{eq:cost-law}.

Once $G$ reaches the largest $m_c$, each state ciphertext contributes one group and the sum becomes $C_{live}$. The packed count therefore stays the same as the context grows until the context needs another state ciphertext, and then it rises by the cost of that ciphertext. Operands that share a bootstrap also share the softmax denominator, and the unpacked circuit computes the same denominator, so packing preserves the value. We use the largest $m_c$ as the group width.

\subsection{Multiple GPUs}\label{sec:device-axis}

The optimizations above lower the bootstrap count of a step, and the bootstraps that remain run one after another on one GPU. The third optimization spreads them over several GPUs, which shortens the critical path of a step and leaves its bootstrap count unchanged. The schedule determines which bootstraps are independent, and a barrier is a set of mutually independent bootstraps that the step completes before proceeding. All operands of a pair go through the same operation sequence, so the circuit executes each site for all pairs or omits it for all pairs. A barrier therefore contains one site across all pairs. The width of a barrier is the number of bootstraps in it. The schedule fixes the number of barriers and the width of each before the run, and the widths are set by the number of state ciphertexts $C$.

Several devices can share the bootstraps of one barrier, or the other devices, the workers, can take a fixed set of bootstraps from the device that runs the step, the master. Sharing a barrier of width $w$ across $P$ devices leaves $\lceil w/P \rceil$ bootstraps on the critical path. Offloading assigns bootstraps whose operands lie in the reserved ciphertexts to worker devices, and the master device continues with the rest. Both placements follow from the schedule. Each barrier ends in a synchronization, after which the step continues on the master device.

A wider model holds fewer positions in each ciphertext. The same context then needs more state ciphertexts, and each barrier is wider. The larger models we project later need several devices to hold their state at all, so the value of a second device grows with the model.

\paragraph{Saturation at three devices.} For each barrier width $w$ that occurs in a step of the measured model, $\lceil w/3 \rceil$ equals $\lceil w/4 \rceil$, so a third device is the last that shortens the critical path. A fourth device can remove only the master device's share of work.

\paragraph{Composition with packing.} Packing saves most of the bootstraps that barrier sharing would spread across devices. Offloading takes a fixed set of bootstraps that packing leaves in place. The two gains therefore act on different bootstraps, and the evaluation measures them together.

\begin{figure}[t]
\centering
\includegraphics[width=\columnwidth]{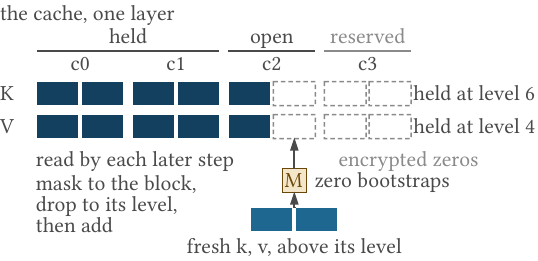}
\caption{The cache on the state layout, one layer. One key and one value ciphertext for each state ciphertext, held at the levels attention consumes. Later steps read the held entries, and the reserved entry holds encrypted zeros.}
\Description{Key and value rows of held, open and reserved ciphertexts of two blocks each with their levels, and a new operand masked and added into the open block.}
\label{fig:cache}
\end{figure}

\subsection{Caching keys and values under encryption}\label{sec:cache}

 The fourth optimization is an encrypted cache of keys and values. Without it, $B(n)$ grows with the context because attention rebuilds the keys and values of all past positions at each step. The keys and values of a position do not change after the step that produces them, so the server keeps them as ciphertexts and later steps read them instead of rebuilding them.

The cache holds one key ciphertext and one value ciphertext for each layer and state ciphertext, laid out like the state with one position per block. Figure~\ref{fig:cache} shows one layer of it on the state layout. Each step writes one key block and one value block in each layer, then reads the remaining blocks. To add a position, the server masks the new key or value to the block of that position, drops it to the level attention reads, and adds it to that key or value ciphertext, using only evaluation keys the run already holds. Each entry starts as an encrypted zero at that level, so attention reads zeros for a state ciphertext the context has just reached. The cached score computation replicates the query across blocks through the rotate-and-add that already replicates the new token's embedding across the blocks, at a cost of one level.

The attention term loses its factor of $n$, and a generated step has a bootstrap count of $B_{cached} = \rho_{att} L C + \mathrm{constant}$.

We hold keys at level six and values at level four, the levels at which the attention path consumes them. The cache also lowers the number of ciphertexts held in memory, because the loop without the cache holds a score and a value ciphertext for all past positions. The cache produces the same values, because a cached key is the ciphertext the loop would otherwise rebuild. All entries stay encrypted, and the cache does not change what the server learns.

\subsection{Implementation}
\label{sec:stack}
We implemented the decode loop and its five stages, the bootstrap schedule, retrieval and write-back, the key and value cache, the optimization configurations and the driver on NEXUS's released evaluator over Phantom \cite{nexus, phantom}. Phantom allocates device memory, launches the kernels and performs the ring arithmetic, the key switching and the bootstrapping. The placement rule is one call at each site, which compares the operand's level with the stage threshold and bootstraps below it, and the loop therefore keeps no schedule table. GELU is a Chebyshev polynomial of degree 47, since a power basis of that degree overflows CKKS precision. Stage markers time each stage under the profiler, and the run without the profiler carries none of them. We will make the code publicly available.
\section{Evaluation}

\label{sec:evaluation}

In this section we compare each step's bootstrap count with Equation~\ref{eq:cost-law}, measure the bootstraps and runtime that each optimization saves, measure the step on one to four devices, and report the agreement of the decoded tokens with the plaintext model. GPT-2 small and nanoGPT run end to end. We estimate the cost of Llama-3-8B from measured operator costs and project larger models from the formula.

\subsection{Experimental setup}
\label{sec:implementation}

All runs we report pass a check that each of their bootstraps returned the number of levels the placement rule assumed. In each run the program also evaluates the plaintext model on the same weights in double precision, in the same process, and a second, independently written implementation checks that evaluation. We run each configuration twice, once without the profiler and once under it. The profiled run counts the bootstraps independently and its count agrees with the program's own, and profiling lengthens the runtime by about 1\%, and all wall clocks we report come from the run without the profiler.

\paragraph{The parameter set.}\label{sec:parameters} All measurements use the parameter set of Section~\ref{sec:background}. At ring degree $2^{15}$ the 128-bit security budget leaves the circuit without a usable level after a CKKS bootstrap \cite{hestandard, curtisplayer, ckksboot}. Splitting the special modulus in two halves the size of each evaluation key and the number of key-switching digits \cite{hanki}, and the whole key set stays resident on one device for the run. The full parameter set is in Appendix~\ref{sec:supp-card}.

\begin{table}[t]
\centering
\small
\caption{The constants of Eqn~\ref{eq:cost-law} for the two measured models and the five projected ones, under the CutMax argmax. Each model reads $B(n) = sCn + cC + k$ with $s = \rho_{att} L$ and $C = \lceil n_{\max}/b \rceil$, and $k$ is given for the prompt step and the generated step. The rows below the rule are derived from each model's circuit.}
\label{tab:constants}
\setlength{\tabcolsep}{3pt}
\begin{tabular}{@{}lrrrrrr@{}}
\toprule
 & $L$, $K$, $V_{ct}$ & $b$ & $\rho_{att}$, $\rho_{lay}$, $\rho_{ffn}$ & $s$ & $c$ & $k$ \\
\midrule
nanoGPT & 6, 3, 1 & 32 & 8, 21, 10 & \LawSlope & \LawColConst & \LawPromptConst, \LawGenConst \\
GPT-2 small & \GptTwoLayers, \GptTwoChunks, \GptTwoLogitCts & 16 & 8, 21, 10 & \GptTwoLawSlopeCol & \GptTwoLawColConst & \GptTwoLawPromptConst, \GptTwoLawGenConst \\
\midrule
Llama-3-8B & \LlamaEightLayers, \LlamaEightChunks, \LlamaEightLogitCts & \LlamaEightPositions & 8, 19, 10 & \LlamaEightLawSlope & \LlamaEightLawColConst & \LlamaEightLawPromptConst, \LlamaEightLawGenConst \\
Mistral-7B & \MistralLayers, \MistralChunks, \MistralLogitCts & \MistralPositions & 8, 19, 10 & \MistralLawSlope & \MistralLawColConst & \MistralLawPromptConst, \MistralLawGenConst \\
Qwen3-8B & \QwenEightLayers, \QwenEightChunks, \QwenEightLogitCts & \QwenEightPositions & 8, 19, 10 & \QwenEightLawSlope & \QwenEightLawColConst & \QwenEightLawPromptConst, \QwenEightLawGenConst \\
Llama-3-70B & \LlamaSeventyLayers, \LlamaSeventyChunks, \LlamaSeventyLogitCts & \LlamaSeventyPositions & 8, 19, 10 & \LlamaSeventyLawSlope & \LlamaSeventyLawColConst & \LlamaSeventyLawPromptConst, \LlamaSeventyLawGenConst \\
Qwen2.5-72B & \QwenSeventyTwoLayers, \QwenSeventyTwoChunks, \QwenSeventyTwoLogitCts & \QwenSeventyTwoPositions & 8, 19, 10 & \QwenSeventyTwoLawSlope & \QwenSeventyTwoLawColConst & \QwenSeventyTwoLawPromptConst, \QwenSeventyTwoLawGenConst \\
\bottomrule
\end{tabular}
\end{table}
\paragraph{Models, metrics and checks.}\label{sec:method}\label{sec:instances} We report three models. GPT-2 small \cite{gpt2} runs the whole loop, Llama-3-8B \cite{llama3} is derived from operator costs measured on it, and nanoGPT carries the sweep over context lengths, the device measurements and the prompt-length check. nanoGPT is the character-level model of the same architecture \cite{nanogpt} with a 65-token vocabulary, and we report its agreement with the plaintext model. We project four more models with Equation~\ref{eq:cost-law} at the measured bootstrap cost, Mistral-7B, Qwen3-8B, Llama-3-70B and Qwen2.5-72B \cite{mistralseven, qwenthree, llama3, qwen25}, dense decoder-only transformers whose layer count, width, feed-forward width, heads and vocabulary we read from their reports. A projected runtime counts only the bootstraps of a step and is a lower bound on it.

All measurements in this paper run on one machine with four NVIDIA H100 GPUs of 64 GB each. Unless stated otherwise, we use one device, and each run uses our parameters and random weights for each model's vocabulary. We count the bootstraps of each step. The plaintext model on the same weights is the reference for correctness. Token agreement is whether the selected token equals the plaintext model's token at that step, and we report the index of the first step at which they differ. Alignment is the cosine between the decrypted final hidden state and the plaintext model's on the same token history, and it is 1 at exact agreement. CKKS latency is independent of the weight values because the operations, their order and each operand's level are fixed before the run. The correctness check is exact at any weights. Alignment does not determine token agreement, since one measured step returned the right token at alignment 0.754 and another returned the wrong token at 0.9997.

\paragraph{Determinism.} Decoding is greedy, because selection returns the argmax of the logits at each step. One weight file serves all runs, and its generator reproduces that file byte for byte. We repeated the run at \HorizonReplicatedLengths prompt lengths, and each repetition returned the same tokens as the first run. A disagreement with the plaintext model is therefore the encrypted circuit's approximation error. We measure each optimization by running the same step with and without it in one process. Two runs of one configuration return the same counts and tokens and differ only in the encryption randomness, so we compare counts first, tokens second, and decrypted values against that spread.

\subsection{The cost of a token}\label{sec:token-cost}

\begin{figure}[t] \centering \includegraphics{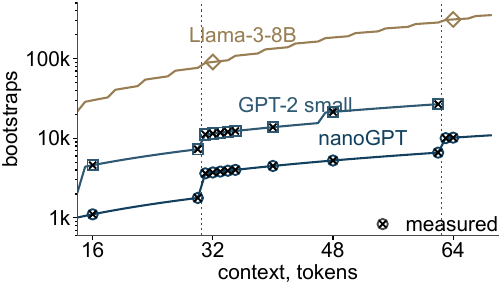} \caption{Bootstraps per prompt step against context, on one device. Each line is the closed form of one model, and a cross inside a ring is a measurement that falls on the prediction. The Llama-3-8B rings are empty because we did not run a step on that model.} \Description{One panel over a context axis, drawing bootstraps per prompt step on a logarithmic count axis for three model architectures, each in its own colour, two of them carrying measurements.} \label{fig:law} \end{figure}

We first compare the bootstrap counts Equation~\ref{eq:cost-law} predicts with the counts we measure on nanoGPT and GPT-2 small and derive for Llama-3-8B (Table~\ref{tab:lawcheck}, Figure~\ref{fig:law}), and then convert counts into runtime with the measured cost of one bootstrap. Equation~\ref{eq:cost-law} specifies the bootstrap count for a run, and Table~\ref{tab:constants} gives its constants for the two measured models. At $C = 2$, it predicts \PromptStepBoots bootstraps for the prompt step of a 32-position nanoGPT context and \GenStepBoots for the following generated step. When the write-back of a step is the first write into a reserved state ciphertext, the constant of that generated step is one lower. Two generated steps at the same $n$ can therefore have different counts. Table~\ref{tab:lawcheck} reports the prediction and measurement for each model.

\begin{table}[t]
\centering
\small
\caption{Bootstraps per step, predicted by Eqn~\ref{eq:cost-law} and
measured, at a context of 32 for the prompt step and 33 for the generated
step. At GPT-2 small the measured count is three bootstraps more than the formula predicts, the bootstraps of the selection stage at its vocabulary. The Llama-3-8B cells marked $\ddagger$ are derived, and the number in parentheses is the measured runtime of that step in seconds.}
\label{tab:lawcheck}
\setlength{\tabcolsep}{3pt}
\begin{tabular}{@{}lrrrr@{}}
\toprule
& \multicolumn{2}{c}{Prompt step} & \multicolumn{2}{c@{}}{Generated step} \\
\cmidrule(lr){2-3}\cmidrule(l){4-5}
& predicted & measured & predicted & measured \\
\midrule
nanoGPT & \PromptStepBoots{} & \PromptStepBoots{} (\StockPromptStepSec{}) & \GenStepBoots{} & \GenStepBoots{} (\StockSecPerGenStep{}) \\
GPT-2 small & \GptTwoLawPromptBoots{} & \GptTwoMeasPromptBoots{} (\GptTwoStockPromptStepSec{}) & \GptTwoLawGenBoots{} & \GptTwoMeasGenBoots{} (\GptTwoStockSecPerGenStep{}) \\
Llama-3-8B & \LlamaPromptBoots{} & \LlamaPromptBoots{}$^\ddagger$ & \LlamaGenBoots{} & \LlamaGenBoots{}$^\ddagger$ \\
\bottomrule
\end{tabular}
\end{table}

Equation~\ref{eq:cost-law} is exact at every step we measured, \GptTwoLawReadings measurements at \GptTwoLawContexts contexts on GPT-2 small (Figure~\ref{fig:law}) and each step of nanoGPT's own sweep. Each added context token costs \GptTwoLawSlopeCol bootstraps for each reserved ciphertext, the slope $s$ of Table~\ref{tab:constants}, which gives \GptTwoTokenCostTwoCol at two ciphertexts and \GptTwoTokenCostThreeCol at three. When the context reaches a third ciphertext, the slope grows by \GptTwoLawSlopeCol and the count grows by the per-ciphertext constant $c$ of Table~\ref{tab:constants}. The sweep stops at four reserved ciphertexts because five exceed one device's memory.

For Llama-3-8B, whose step exceeds the memory of one device, the closed form and a level-by-level simulation of the circuit with our thirteen usable levels return the same counts. Table~\ref{tab:lawcheck} lists them beside the two measured models.

A bootstrap count becomes a wall clock through the duration of one bootstrap, which is \RefreshPriceMs ms on an idle device with our parameter set and varies by at most 4.2\% across different runs.

\subsection{Optimization results}\label{sec:optimization-results}

\begin{figure}[t]
\centering
\includegraphics{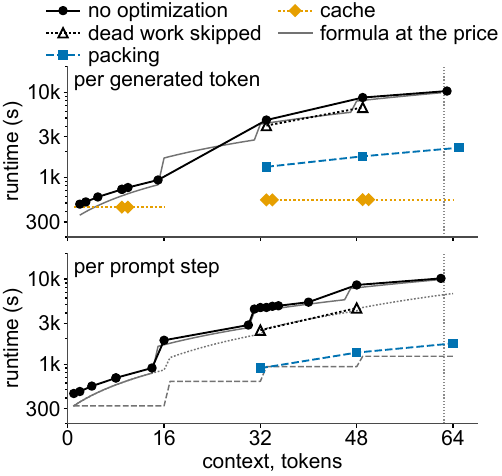}
\caption{Runtime per generated token (top) and per prompt step (bottom) against context at GPT-2 small, on a logarithmic runtime axis. Points
are measured. Each grey line is a configuration's bootstrap count at the measured
cost, and the cached line is the measured step at each reserved ciphertext count. The dotted vertical line marks the fifth reserved ciphertext, beyond which the naive and the cached configurations exceed the memory of one device.}
\Description{Two stacked plots of seconds against context, the generated
step above and the prompt step below, one axis a panel and one legend over
both.}
\label{fig:seconds}
\end{figure}

We measure what each optimization saves on GPT-2 small, over a two-step run and at the prompt step (Table~\ref{tab:shares}), for one generated step under each configuration on the three models (Table~\ref{tab:results}), and as runtime against context (Figure~\ref{fig:seconds}). Table~\ref{tab:shares} gives the fraction of a step's bootstraps and of its wall clock that each optimization saves, over one prompt step and one generated step together and at the prompt step alone. The two fractions differ because a step spends part of its time outside bootstrapping.

\begin{table}[t]
\centering
\small
\caption{The share of a step each optimization saves at GPT-2 small.
The two-step values come from a run of two steps at the shortest context we ran, one process for each pair, and count the bootstraps and the wall clock
of both steps. The prompt step values come from the runs of Figure~\ref{fig:seconds} at two contexts and read the wall clock. The rows are cumulative. The packing row measures packing with dead work skipping applied.}
\label{tab:shares}
\setlength{\tabcolsep}{4pt}
\begin{tabular}{@{}lrrrr@{}}
\toprule
& \multicolumn{2}{c}{Two steps} & \multicolumn{2}{c@{}}{Prompt step} \\
\cmidrule(lr){2-3}\cmidrule(l){4-5}
& bootstraps & wall clock & context 32 & context 48 \\
\midrule
Dead work & \GptTwoDeadCountPct\% & \GptTwoDeadWallPct\% & \TtftDeadSharePctLow\% & \TtftDeadSharePctHigh\% \\
+ packing & \GptTwoPackCountPct\% & \GptTwoPackWallPct\% & \TtftPackSharePctLow\% & \TtftPackSharePctHigh\% \\
\bottomrule
\end{tabular}
\end{table}

 At the prompt step the reservation saves more than causality does, and together they save the sum of the two. At the generated step the reservation leaves the count unchanged, because the context has reached its last reserved ciphertext, and causality saves the same amount alone and with the reservation.

\begin{figure}[t]
\centering
\includegraphics{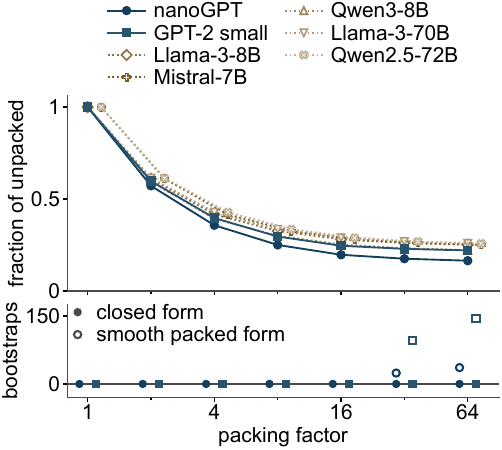}
\caption{Bootstraps per prompt step at context 48, as a fraction of the count at group width $G = 1$ (top). The two blue models are measured and the five tan ones come from the closed form. Mistral-7B shares the layer count and positions per ciphertext of Llama-3-8B, and Qwen2.5-72B those of Llama-3-70B, so each pair's curves coincide and the second of each pair is drawn slightly to the right. The measured count minus the closed
form is zero at every width (bottom), and the smooth form without the
ceiling falls under the measurement at the two widest.}
\Description{Two stacked plots against the packing factor. The upper plot
draws seven model architectures falling from one, two of them measured and five priced from the closed form. The lower plot draws the measured count minus
the closed form, zero everywhere, and minus the smooth packed form, above
zero at the two widest groups.}
\label{fig:packing}
\end{figure}

The packed closed form, Equation~\ref{eq:packed}, reproduces the measured count at each of the \GptTwoPackedPromptReadings packed prompt steps across three context lengths and seven group widths (Figure~\ref{fig:packing}). At the generated step, it reproduces \GptTwoPackedGenHits of \GptTwoPackedGenReadings. At group widths of \GptTwoPackedMissWidth or narrower, the form predicts \GptTwoPackedGenMissLow to \GptTwoPackedGenMissHigh bootstraps fewer than we measure. At the widest group, packing saves 85\% or more of the bootstraps at all contexts we ran.

\begin{table}[t]
\centering
\small
\caption{One generated step at a context of 33 on one device. Each configuration adds its optimization to the one before it. The rows above the rule are measured, except the dead-work runtime of nanoGPT, which is that count times the measured bootstrap cost. The rows below the rule are derived: their counts are the closed form and their runtime is the count times the measured bootstrap cost, a lower bound on the step.}
\label{tab:results}
\setlength{\tabcolsep}{5pt}
\begin{tabular}{@{}lrrrr@{}}
\toprule
\textbf{Bootstraps}& None & + dead work & + packing & + cache \\
\midrule
nanoGPT & \GenStepBoots{} & \DeadWorkGenBoots{} & \PackedGenBoots{} & \CachedStepBoots{} \\
GPT-2 small & \GptTwoMeasGenBoots{} & \GptTwoDeadGenBoots{} & \GptTwoPackedGenBoots{} & \GptTwoCachedStepBoots{} \\
\midrule
Llama-3-8B & \LlamaGenBoots{} & \LlamaDeadGenBoots{} & \LlamaPackedGenBoots{} & \LlamaCachedBoots{} \\
Mistral-7B & \MistralGenBoots{} & \MistralDeadGenBoots{} & \MistralPackedGenBoots{} & \MistralCachedBoots{} \\
Qwen3-8B & \QwenEightGenBoots{} & \QwenEightDeadGenBoots{} & \QwenEightPackedGenBoots{} & \QwenEightCachedBoots{} \\
Llama-3-70B & \LlamaSeventyGenBoots{} & \LlamaSeventyDeadGenBoots{} & \LlamaSeventyPackedGenBoots{} & \LlamaSeventyCachedBoots{} \\
Qwen2.5-72B & \QwenSeventyTwoGenBoots{} & \QwenSeventyTwoDeadGenBoots{} & \QwenSeventyTwoPackedGenBoots{} & \QwenSeventyTwoCachedBoots{} \\
\bottomrule
\addlinespace

\textbf{Runtime (s)}& None & + dead work & + packing & + cache \\
\midrule
nanoGPT & \StockSecPerGenStep{} & \DeadWorkGenSecDerived{} & \PackedSecPerGenStep{} & \CachedSecPerGenStep{} \\
GPT-2 small & \GptTwoStockSecPerGenStep{} & \GptTwoDeadSecPerGenStep{} & \GptTwoPackedSecPerGenStep{} & \GptTwoCachedSecPerGenStep{} \\
\midrule
Llama-3-8B & \ProjLlamaEightGenSec{} & \LlamaDeadGenSec{} & \LlamaPackedGenSec{} & \ProjLlamaEightCachedSec{} \\
Mistral-7B & \MistralGenSec{} & \MistralDeadGenSec{} & \MistralPackedGenSec{} & \MistralCachedSec{} \\
Qwen3-8B & \QwenEightGenSec{} & \QwenEightDeadGenSec{} & \QwenEightPackedGenSec{} & \QwenEightCachedSec{} \\
Llama-3-70B & \LlamaSeventyGenSec{} & \LlamaSeventyDeadGenSec{} & \LlamaSeventyPackedGenSec{} & \LlamaSeventyCachedSec{} \\
Qwen2.5-72B & \QwenSeventyTwoGenSec{} & \QwenSeventyTwoDeadGenSec{} & \QwenSeventyTwoPackedGenSec{} & \QwenSeventyTwoCachedSec{} \\
\bottomrule
\end{tabular}
\end{table}

Table~\ref{tab:results} lists one generated step under each configuration for the three models.
Of the four optimizations, the cache saves the most bootstraps. The cached form of Equation~\ref{eq:cost-law} gives the measured count at every cached step. Across the two cached steps of a run, the cached-step count is constant in context and increases by one ciphertext's share at the boundary to \GptTwoCachedStepBootsFour bootstraps at four reserved ciphertexts. The prompt step costs \GptTwoPromptStepSec seconds because it carries the whole context without the cache. On one device the first token arrives after \GptTwoPromptStepHours hours, and the loop then generates \GptTwoTokensPerHour tokens an hour. We compare the cached loop against the naive implementation in one process with one key set. Tokens are identical at all steps, and both configurations agree with the plaintext model. The two configurations are the None and Cache entries of the GPT-2 small row of Table~\ref{tab:results}.

At both models each cached step reads the cache without bootstrapping it, and each added key or value arrives five levels above the level at which attention reads it.

We measure prompt-step cost on the same runs. Figure~\ref{fig:seconds}, bottom, plots time to first token, the prompt step's wall clock, against prompt length at GPT-2 small. The prompt step's wall clock grows linearly with prompt length at a fixed reserved ciphertext count and jumps at each ciphertext boundary. The count of Equation~\ref{eq:cost-law} multiplied by the measured bootstrap cost falls \TtftPriceGapPctLow to \TtftPriceGapPctHigh\% below the measured wall clock. The cache leaves the prompt step unchanged, and with and without it in one process the prompt step's wall clock differs by less than the variation between two runs of one configuration.

\paragraph{GPT-2 small and nanoGPT.} Skipping dead work and packing cut a run of two steps on GPT-2 small by a factor of \GptTwoTwoStepFactor in one process with one context and one key set, and the run costs \GptTwoSecPerToken seconds a token within one device's memory. On nanoGPT the four optimizations together take a generated step from \GenStepBoots to \CachedStepBoots bootstraps and from \StockSecPerGenStep to \CachedSecPerGenStep seconds (Table~\ref{tab:results}). The two models differ in width and in vocabulary. Enlarging the vocabulary from 65 tokens to GPT-2's adds 18 bootstraps to a step, under 0.5\% of the step, at both step types and both models, and costs memory for the logit and retrieval ciphertexts.

\paragraph{Llama-3-8B.} We measured each operator of Llama-3-8B under our parameter set and use the bootstrap cost measured on the same GPU. The counts of Table~\ref{tab:lawcheck} multiplied by the measured bootstrap cost give the bootstrap time of a step. Table~\ref{tab:results} gives the four configurations on this model on the same basis.

\subsection{Multi-device setting}\label{sec:multi-device}

We measure the device axis on nanoGPT, whose generated step with dead work skipped and packing at the widest group takes \DevGridStepOne seconds on one device. Table~\ref{tab:device-axis} gives, at one to four devices, the number of bootstraps on the critical path, the step time our model predicts from that number, and the measured step time. The number of offloaded bootstraps on the critical path falls up to the third device, and a fourth device does not shorten it further. The predicted step time lies within \DevModelRunOnePct\% of run 1 at each device count and within \DevModelRunTwoPct\% of run 2.

\begin{table}[t]
\caption{The device axis on nanoGPT, one generated step at a context of 33
with dead work skipped and packing at the widest group. The critical path is
the number of offloaded bootstraps the step waits for. The model adds that path at the measured bootstrap cost to the master device's \CeilMasterS seconds. Run 2 repeats run 1 on a clean node, and the
speedup is against one device within each run.}
\label{tab:device-axis}
\centering
\begin{tabular}{@{}ccccccc@{}}
\toprule
& Critical path & \multicolumn{3}{c}{Generated step (s)} & \multicolumn{2}{c}{Speedup} \\
\cmidrule(lr){3-5} \cmidrule(l){6-7}
\#GPUs & (bootstraps) & model & run1 & run2 & run1 & run2 \\
\midrule
1 & 227 & 342 & 337 & 336 & 1.00 & 1.00 \\
2 & 114 & 300 & 300 & 295 & 1.12 & 1.14 \\
3 & 102 & 296 & 300 & 306 & 1.12 & 1.10 \\
4 & 102 & 296 & 292 & 297 & 1.15 & 1.13 \\
\bottomrule
\end{tabular}
\end{table}

Two and four devices cut the generated step by \DevStepCutLowPct to \DevStepCutHighPct\% against one device in both runs. Three devices cut it by \DevThreeCutGridPct\% in run 1 and \DevThreeCutRepPct\% in run 2. In each configuration of run 1 the bootstrap count is the same at all device counts.

Two terms set the lower bound of the generated step. The master device's share of a generated step is \CeilMasterS seconds at each device count, and the offloaded bootstraps take at least one bootstrap per barrier. The generated step therefore cannot fall below about \CeilStepFloorS seconds, and one device is at most a factor of \CeilFactor above that floor. The measured steps at two to four devices stay \CeilAboveFloorLowPct to \CeilAboveFloorHighPct\% above it.

The same runs also show how several devices interact with packing. Worker devices take a fixed set of bootstraps off the master device, and packing does not change that set. Packing shortens the rest of the step, so the offloaded set is a larger share of a packed step than of an unpacked one. At the same device count, a packed step speeds up by a factor of 1.09 and an unpacked step by 1.02. The two optimizations act on different bootstraps.

\subsection{Correctness}\label{sec:correctness}

We verify correctness against the plaintext model on the same weights, by token agreement and by the alignment of the decrypted final hidden state. We then confirm that a step's cost does not depend on the token it returns, and state the security level of the parameter set. Each stage of the loop evaluates a polynomial approximation of the function it replaces. The encrypted loop follows the plaintext model closely and can differ from it at single tokens. At each prompt length we ran on nanoGPT, the alignment of the decrypted final hidden state with the plaintext model's stays above \ALIGNMENTFLOOR. Where the plaintext model's two highest logits are close, encrypted selection can return the second of them.

\paragraph{The context boundary.} We ran one prompt of each length from 40 to \HorizonExactAgainTo{} tokens on nanoGPT, generated two tokens from each and compared every token with the plaintext model's. The loop returns the plaintext model's tokens at every length except \HorizonDivergent, where its first token differs, and it is exact again at the longer prompts. Exactness therefore does not degrade with the context inside this range. At the diverging step the selection stage is less certain of its winner than at the exact lengths, while the hidden state stays aligned with the plaintext model's. Appendix~\ref{sec:supp-correctness} localizes the failure to one operation, the first inverse of the attention softmax, where the additive error of one bootstrap consumes an operand that shrinks with the context, so the boundary sits at one site of the circuit.

\paragraph{Cost against the token returned.} A step costs the same whatever token it returns. We ran three prompts under the same configuration, and their bootstrap counts were identical at all eight steps, from 3716 to 4390. Their wall clocks differed by at most 1.827\% within the 2.32\% spread between repeated runs. The prompt whose selected token differs from the plaintext model's has the counts the formula predicts. The same holds inside one operator, since the softmax microbenchmark's time changed by 0.1\% when its input changed.

We repeat the check on GPT-2 small with its full vocabulary, and the encrypted circuit returns the plaintext model's token at the first step, as it does on nanoGPT. The disagreement rate does not grow with the context.

The parameter set of each run also fixes the security level. The lattice estimator \cite{lweestimator} certifies our parameters at not less than \SecurityBits bits of classical security. The cheapest attack the estimator finds is a hybrid attack on bounded distance decoding with a meet-in-the-middle step, and it costs $2^{\SecurityAttackLogRop}$ operations under the estimator's default reduction cost model. We claim classical security only. Appendix~\ref{sec:supp-security} proves the statement of Section~\ref{sec:threat-model}. A simulator that receives only the public description of the request, the model, the packing, the prompt length and the token count, encrypts zeros in place of the prompt's ciphertexts, and a hybrid argument over those ciphertexts reduces any distinguisher of the two views to the IND-CPA security of CKKS with the evaluation keys published, with a loss linear in the number of prompt ciphertexts.

\begin{table}[t]
\centering
\small
\caption{Four operations of our loop beside NEXUS's, at NEXUS's own parameters. For GELU, LayerNorm and softmax we run NEXUS's released binary and ours on the same device and report medians over \CmpCalls calls. For the matrix multiplication we report \CmpMatmulReps repetitions on our H100 next to the runtime NEXUS published for an A100, so the two numbers are not comparable.}
\label{tab:cmp}
\setlength{\tabcolsep}{5pt}
\begin{tabular}{@{}lrr@{}}
\toprule
Operation & Ours & NEXUS \\
\midrule
GELU & \CmpGeluMs{} ms & \CmpGeluNexusMs{} ms \\
LayerNorm & \CmpLnMs{} ms & \CmpLnNexusMs{} ms \\
Softmax & \CmpSoftmaxMs{} ms & \CmpSoftmaxNexusMs{} ms \\
Matrix multiplication/column & \CmpMatmulSecPerCol{} s  & \CmpNexusMatmulSecPerCol{} s  \\
\bottomrule
\end{tabular}
\end{table}

\subsection{Comparison}\label{sec:comparison}

Our end-to-end baseline is the naive implementation of the loop, the first row of Table~\ref{tab:results}. Table~\ref{tab:cmp} compares our loop's operators with NEXUS's readings of the same operators at NEXUS's parameters, which differ from ours. Our GELU runs 2\% slower and our LayerNorm 3\% faster than NEXUS's on the same device, and both differences lie within the variation between two runs of one configuration. NEXUS measured its matrix multiplication on an A100, so the table lists the two numbers side by side and we do not compare them. The decoded output of each operator passes its check against the reference computation.

We also derive the cost of the interactive loop of Section~\ref{sec:placement-problem}. Each assumption in that derivation favours the interactive loop. It saves \InteractiveSavingPctRounded\% of server time per token, and the client must stay online for \InteractiveClientHours hours to receive sixteen tokens. At Llama-3-8B, the same derivation saves \LlamaInteractiveRemoved bootstraps from the cached generated step of \LlamaCachedBoots, a saving of \LlamaInteractiveSavingPct\%, for \LlamaInteractiveClientHours hours of client presence at sixteen tokens.

\subsection{Scaling in model and context size}

\label{sec:discussion}

\begin{figure}[t]
\centering
\includegraphics{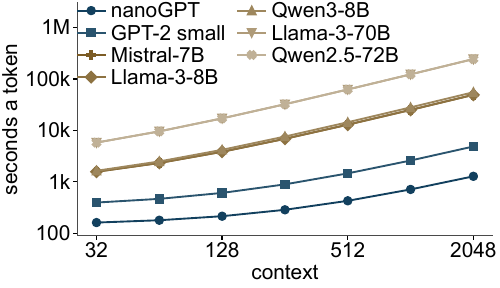}
\caption{The cached generated step against the context for \ProjSlateShapeCount models, the closed form at \RefreshPriceMs~ms a bootstrap. Each curve counts the bootstraps of a step and is a lower bound on the step.}
\Description{Seven curves rising with the context on logarithmic axes, each in its own colour, two of them lying together.}
\label{fig:projbody}
\end{figure}

The rows below the rule of Table~\ref{tab:results} give the four configurations for the four projected models at a context of 33, and this subsection extends the projection to longer contexts, larger models and more devices. We estimate the cached generated step of \ProjSlateShapeCount dense models from a context of 32 to \ProjContextMax with Equation~\ref{eq:cost-law} and the measured bootstrap cost (Figure~\ref{fig:projbody}). Over that range the cached token cost grows by a factor of \ProjLlamaSeventyGrowth at Llama-3-70B and \ProjMistralGrowth at Mistral-7B.

\begin{figure}[t]
\centering
\includegraphics{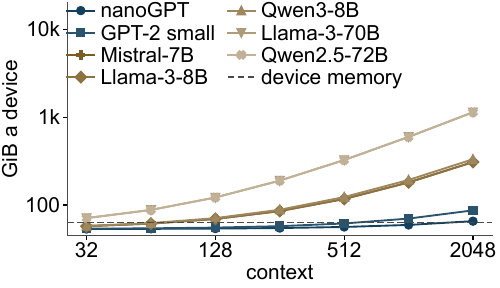}
\caption{Memory footprint of a run on one device against the context for seven models, with the cache, from unit prices measured on this parameter set. The dashed line is the memory of one device.}
\Description{Seven memory curves rising with the context on logarithmic axes. A dashed horizontal line marks one device, and the two largest models start above it.}
\label{fig:memory}
\end{figure}

Figure~\ref{fig:memory} gives the memory one device needs against the context for the seven models. The evaluation keys take the largest share, \ProjKeySetGiB{} GiB, and the rest is the state, the working set of a step and the cache. One device holds nanoGPT to a context of \ProjFitsNanoCtx, GPT-2 small to \ProjFitsGptTwoCtx{} and the 8B models to \ProjFitsEightBCtx, and the 70B models need more than one device at every context. On the two models we measure, the estimate is within \ProjMemErrPct{} per cent of the footprint. A projected model may need several devices to hold its state, and we split its layers across devices in a pipeline, as plaintext serving does for a model larger than one device. The split divides the cache among the devices, leaves the bootstraps and the critical path unchanged, adds one state transfer at each layer boundary, and raises throughput by the device count over independent requests. Within one device's share, a wider model holds fewer positions in each ciphertext, so its barriers are wider and offloading removes a larger share of the step than on nanoGPT. Models in use today add modules and methods beyond the dense transformer, such as mixtures of experts, linear attention and sparse attention, which bring new nonlinear operators and change the terms of the formula. The methodology stays as it is, since the placement rule fixes the bootstraps of the new operators before the run, and the count of a step follows from the same ledger of sites.

\paragraph{Scope.} By design the framework does not depend on pretraining. Each run uses random weights, since the schedule and the bootstrap count depend on the architecture and not on the weight values, and we make no claim about language quality. The number of tokens to generate is fixed before the run, because the server cannot see whether the model has produced an end-of-sequence token. We measure two models and derive the rest from two untested assumptions. Where we have both a projection and a measurement, the projection is \ProjShortfallPctLow to \ProjShortfallPctHigh\% below the measurement, and the gap is widest at the cached step. We use one device for the operator comparison.

\section{Related work}

\label{sec:related}

\paragraph{Non-interactive inference.} Prior systems optimize private transformer forward passes under CKKS, and NEXUS made this setting non-interactive \cite{nexus}. Its client sends one encrypted input and receives one encrypted output, and the server computes everything between them, the final selection vector included, under encryption. NEXUS's final output is an encrypted one-hot vector, the same object our selection stage produces at every step. Because no published system selects a token under encryption and feeds it back into the encrypted state, we derive the nearest alternative, an interactive loop, and compare against it in the evaluation. A per-token ratio against a forward-pass system would credit our loop for the forward pass, which is their work.

\paragraph{Scale-out inference and encrypted caches.} Cerium is a scale-out CKKS framework for large models across GPUs, and its reported task covers only the first token \cite{cerium}. Cerium reports its bootstrap time from a standalone benchmark, so it cannot be set beside ours, which we time inside the decode loop. Cachemir is the closest work to our cache, because it packs an encrypted key and value cache for CKKS transformer inference and reports end-to-end latency at fixed sequence lengths \cite{cachemir}. Cachemir reaches 128 bits of security with a sparse secret through sparse-secret encapsulation.

Cachemir's decoding step takes the most recent token as input and appends its key vector to the cache. Cachemir reports MRPC accuracy and F1 and Wikitext-2 perplexity for that pipeline. These metrics therefore measure a cache that a supplied token sequence fills.

\paragraph{Forward-pass building blocks.} Matrix multiplication kernels \cite{jkls, thor, tricycle} and packing layouts \cite{moai, stip, ellmo} cover the linear stages. Polynomial approximations \cite{powerformer, leaf, chakraborty22} and encryption-friendly architectures \cite{encfriendly} cover the nonlinear stages. ROSETTA extends Cachemir with TFHE lookup tables for the nonlinear operators of a decode step, and a cost model assigns each operator to TFHE or to CKKS \cite{rosetta}. Libraries \cite{openfhe, phantom, fideslib, cheddar}, GPU kernels \cite{zephyr}, and GPU systems for whole models \cite{encllm, sylph} execute these stages, and hardware accelerators, real or simulated, scale the same stages out \cite{cinnamon, craterlake, ark}. Faster operators from this work would also shorten our step, because we use these operators and change the computation around them. Encrypted inference for other model families uses the packing and bootstrap machinery and stops at the forward pass \cite{helrm}. Surveys \cite{sok935, sok047, survey105} and a tutorial \cite{edbt} map the area. The tutorial states that ``fully end-to-end private text generation under HE remains an open challenge'' \cite{edbt}. Hyperion's authors write that they ``plan to integrate this into a larger private LLM system and enable full private LLM token generation under homomorphic encryption'' \cite{hyperion}.

\paragraph{Hybrid protocols.} A separate body of work reaches transformer inference \cite{iron, bolt, bumblebee} and generative models \cite{ciphergpt, neverdecode} by combining homomorphic encryption with multi-party computation, or by secret sharing alone \cite{puma}. The client stays online for every nonlinearity in these protocols, so their threat model differs from ours.

\paragraph{Bootstrap placement.} The systems above run a forward pass, and another body of work places program bootstraps. DaCapo \cite{dacapo}, Orion \cite{orion}, ReSBM \cite{resbm}, Orbit \cite{orbit}, and HEIR \cite{heir} search for a bootstrap schedule over the graph of a program, and HALO \cite{halo} extends that search to a loop with one repeated body. These systems find a cheaper placement inside a fixed graph. Our rule fixes the bootstrap count of a growing body before the run, so the two act on different terms of one step.

When a step is spread across several devices, the schedule must also decide which device runs each bootstrap. Concurrent work on multi-GPU CKKS transformer inference reports that distributing a key switch across devices at the granularity of its digits is bound by communication \cite{aegis}. We distribute whole bootstraps instead, in the groups the schedule fixes before the run, and the evaluation gives the device count at which the gain stops. Encrypted control also runs a feedback loop over encrypted state, with a controller in place of a language model \cite{kogiso2015, farokhi2017, darup2021}.

\paragraph{Selection under encryption.} Prior work studies token selection under encryption, and we build on it. We use CutMax's comparison-free argmax, which converts an encrypted logit vector into a one-hot vector by repeated sharpening \cite{cutmax}. We add the stage that consumes that one-hot vector inside the decode loop. Three other constructions provide the same primitive, an encrypted argmax that returns a one-hot vector at small class counts \cite{phoenix}, probabilistic selection over encrypted logits \cite{tyche}, and inverse-transform sampling at full vocabulary \cite{hyperion}. We decode greedily, which makes the plaintext model an exact comparator, and the loop is independent of the sampling policy. One paper studies how tokens are ordered when the feedback is encrypted, and it does so in plaintext simulation \cite{rhotsp}. Our correctness results measure that failure class on the encrypted circuit itself.
\section{Conclusion}

\label{sec:conclusion}

We presented \Sys{}, the first system to run autoregressive generation under CKKS with the client offline between the prompt and the output. One placement rule fixes the bootstraps of a step before the run and gives their count as a formula in the context length, which held at every step we measured. Skipping dead work, packing and an encrypted key and value cache reduce a GPT-2 small generated step from \GptTwoMeasGenBoots bootstraps to \GptTwoCachedStepBoots, and a second device cuts the wall clock of a nanoGPT step by a further \DevStepCutLowPct{}\%. Each operator inside the loop is a published one and the loop is where the cost now lies.

\section*{Acknowledgments}
Authors from Ko\c{c} University have received funding from the European
Research Council (ERC) under the European Union's Horizon 2020 research and
innovation programme (grant agreement No 949587). We acknowledge the EuroHPC
Joint Undertaking for awarding access to the MareNostrum5 supercomputer in
Spain, where the experiments reported in this paper were partially performed.

\bibliographystyle{IEEEtran}
\bibliography{refs}

\appendix
\renewcommand{\thesubsection}{\thesection.\arabic{subsection}}
\section{The parameter set}
\label{sec:supp-card}

Table~\ref{tab:supp-card} lists the parameter set of every measurement in the paper. The lattice estimator \cite{lweestimator} certifies it at not less than \SecurityBits{} bits of classical security. The evaluation names the cheapest attack the estimator finds and its cost under the estimator's default reduction cost model. We claim classical security only.

\begin{table}[h]
\centering
\small
\caption{The parameter set of every measurement in the paper.}
\label{tab:supp-card}
\begin{tabular}{@{}ll@{}}
\toprule
quantity & value \\
\midrule
ring degree & $2^{16}$ \\
slots used & $2^{14}$ \\
secret & sparse ternary, $h=192$ \\
ciphertext primes & $51 + 13 \times 46$ bits \\
special primes & $2 \times 51$ bits \\
positions per ciphertext, $b$ & 32 \\
usable levels & 13 \\
bootstrap depth & 14 \\
scale & $2^{46}$ \\
bootstrap primes & $14 \times 51$ bits \\
modulus, $\log QP$ & 1465 bits \\
evaluation keys & 19.7 GiB \\
\bottomrule
\end{tabular}
\end{table}
\section{The security reduction}
\label{sec:supp-security}

The threat model of the paper states that the server's view is simulatable from public values. This appendix defines the view, the leakage and the assumption, states the theorem and proves it.

\paragraph{The request and the functionality.} We write $g$ for the public description of a request, the model architecture, the packing, the prompt length $n_0$ and the number of tokens requested $T$. The packing consists of the reservation $n_{\max}$, the number of positions the run reserves, the positions per ciphertext $b$ and the group width $G$. The parameter set is fixed and public. We write $W$ for the server's input, the weights and the embedding table in plaintext, and $x$ for the client's input, a prompt of $n_0$ tokens. The ideal functionality $\mathcal{F}_g$ takes $x$ from the client and $W$ from the server and runs greedy decoding of the plaintext model with weights $W$ from $x$ for $T$ steps. It returns the $T$ tokens to the client and the leakage $L(g)$, defined below, to the server. The end-of-sequence token is an ordinary token, the run does not stop at it, and the client truncates after decryption, because a stop test would need a decryption. A sampling policy is a different functionality, and the theorem does not cover it.

The protocol computes an approximation of $\mathcal{F}_g$. Its client output is $T$ encrypted one-hot vectors, whose tokens agree with $\mathcal{F}_g$ on the runs the evaluation reports as exact and differ on the run it reports as divergent. We prove no bound on that error. The theorem compares the server's view with a simulation of it. It says nothing about the joint distribution of the view and the client's output, since that would need a correctness bound.

\paragraph{The leakage.} The leakage $L(g)$ is $g$ together with every object the run derives from $g$ by a public function. These objects are the parameter set, the number $c(g)$ of prompt ciphertexts, the sequence of homomorphic operations with the level of every operand, the position and the count of every bootstrap, and the rotation set $K_g$ of the run, which includes the conjugation. These are functions of $g$ alone for two reasons. A level is metadata that only the operation sequence changes. The placement rule places a bootstrap when an operand's level is below the threshold of the stage it enters, and it depends on nothing else. The dead-work tests depend on the context length and the reservation, both of which are in $g$. The prompt occupies $c(g) = \lceil n_{\max}/b \rceil$ state ciphertexts, and every position at or past $n_0$ encodes zero. Because $n_0$ is in $g$, the padding reveals nothing beyond $g$. Through $L(g)$ the server learns $n_0$, $T$ and the schedule they fix.

\paragraph{The keys and the assumption.} The client is honest. It runs the key generation of CKKS and obtains a sparse ternary secret key $s$ and a public key $\mathsf{pk}$. It also generates the evaluation keys $\mathsf{evk}_g$, the relinearization key and one rotation key for every element of $K_g$. Bootstrapping uses keys of this family and no other \cite{ckksboot}, and the key set of the run contains no second secret. Each of these keys is a key-switching key. It consists of one or more RLWE encryptions under $s$ of a public integer multiple of $f(s)$ for one function $f$ in the class
$$\mathcal{K}_g = \{\, s \mapsto s^2 \,\} \cup \{\, s \mapsto \sigma_k(s) \mid k \in K_g \,\},$$
where $\sigma_k$ is the automorphism that the rotation by $k$ applies.

The game $\mathrm{CPA}^{\mathrm{evk}}_g$ is the IND-CPA game of CKKS with one change, in which the challenger hands the adversary $\mathsf{pk}$ and $\mathsf{evk}_g$ together. The adversary submits two plaintext vectors of one state ciphertext and receives the encryption of one of them under $\mathsf{pk}$ at the level and scale of a prompt ciphertext. It outputs a bit, and its advantage is the absolute difference between the probabilities that it outputs one on the two challenges. $\mathrm{Adv}^{\mathrm{evk}}_g(t)$ is the largest advantage of an adversary that runs in time $t$. We assume that $\mathrm{Adv}^{\mathrm{evk}}_g(t)$ is negligible at every $g$ the paper runs. This is a key-dependent-message assumption on RLWE for the class $\mathcal{K}_g$ under a sparse secret, the circular security that every CKKS deployment that multiplies or rotates makes. No reduction from decisional RLWE to it is known. The lattice estimator certifies the RLWE instance without the key-switching keys, and the assumption includes that publishing them does not lower that level.

\paragraph{The view and the two experiments.} The server is honest but curious, and its view consists of its input $W$, its random tape $r$ and the messages it receives. The tape supplies the coins of the encryptions of zero under $\mathsf{pk}$ with which the server initializes the cache entries, and no other operation of the server uses coins. The messages are $\mathsf{pk}$, $\mathsf{evk}_g$ and the prompt ciphertexts $\mathsf{ct}_1, \dots, \mathsf{ct}_{c(g)}$, each the encryption under $\mathsf{pk}$ of one state ciphertext's share of the encoded prompt, so
$$\mathrm{view} = (W, r, \mathsf{pk}, \mathsf{evk}_g, \mathsf{ct}_1, \dots, \mathsf{ct}_{c(g)}).$$
The program $P_g$ that maps the view to the run is public and the server holds no value in the clear. Every branch, ciphertext and bootstrap of the run is therefore a function of the view and of $L(g)$. The measurement program holds the secret key in the same process and decrypts probes for validation. $P_g$ is that program with every probe removed, since no probe adds a homomorphic operation or affects a ciphertext, a branch of $P_g$ or the step count.

The real experiment $\mathrm{Real}(g,x,W)$ runs the client's key generation, encrypts the state ciphertexts of $x$ under $\mathsf{pk}$, samples $r$ and outputs the view. The ideal experiment $\mathrm{Ideal}(g,x,W)$ runs $\mathcal{F}_g$, which hands the server $L(g)$, and outputs $\mathcal{S}(L(g),W)$. The simulator $\mathcal{S}$ receives the server's input and the leakage and never receives $x$. The theorem quantifies over every $g$, $x$ and $W$, so an adversary who chooses the prompt gains nothing by the choice.

\paragraph{Theorem.} Let $t_{\mathrm{enc}}$ be the time of one public-key encryption of one state ciphertext. There is a simulator $\mathcal{S}$ that runs in the time of one key generation plus $c(g)\,t_{\mathrm{enc}}$. For every $g$, every prompt $x$ of length $n_0$, every $W$ and every distinguisher $D$ running in time $t$,
\begin{multline*}
\bigl|\Pr[D(\mathrm{Real}(g,x,W)) = 1] - \Pr[D(\mathrm{Ideal}(g,x,W)) = 1]\bigr| \\
\le c(g)\cdot \mathrm{Adv}^{\mathrm{evk}}_g\bigl(t + c(g)\,t_{\mathrm{enc}}\bigr).
\end{multline*}

\paragraph{Proof.} The simulator $\mathcal{S}(L(g),W)$ runs the client's key generation and obtains a fresh pair $(\mathsf{pk}', \mathsf{evk}'_g)$, with $K_g$ read from $L(g)$. It encrypts the zero vector $c(g)$ times under $\mathsf{pk}'$, samples $r$, and outputs $(W, r, \mathsf{pk}', \mathsf{evk}'_g, \mathsf{ct}'_1, \dots, \mathsf{ct}'_{c(g)})$. It uses $W$ only as a component of the tuple.

For $0 \le i \le c(g)$ let $H_i$ be the distribution of the real view with $\mathsf{ct}_1, \dots, \mathsf{ct}_i$ replaced by encryptions of the zero vector under the real $\mathsf{pk}$. Then $H_0$ is $\mathrm{Real}(g,x,W)$, and $H_{c(g)}$ and $\mathrm{Ideal}(g,x,W)$ are identically distributed. Key generation is independent of $x$, so the real key pair and the fresh one have the same distribution. Both tuples are the same function of a key pair, $c(g)$ encryptions of zero and $r$.

For a fixed $i$ with $1 \le i \le c(g)$ we build an adversary $B_i$ against $\mathrm{CPA}^{\mathrm{evk}}_g$ that has $x$, $W$ and $D$ as fixed internal inputs. $B_i$ receives $(\mathsf{pk}, \mathsf{evk}_g)$, submits as its two messages the plaintext of the $i$th prompt ciphertext and the zero vector, and receives the challenge $\mathsf{ct}^*$. It encrypts the zero vector under $\mathsf{pk}$ for the positions before $i$ and the prompt's own plaintexts for the positions after $i$, and it samples $r$. It hands $D$ the tuple $(W, r, \mathsf{pk}, \mathsf{evk}_g, \mathsf{ct}_1, \dots, \mathsf{ct}_{i-1}, \mathsf{ct}^*, \mathsf{ct}_{i+1}, \dots, \mathsf{ct}_{c(g)})$ and outputs what $D$ outputs. When $\mathsf{ct}^*$ encrypts the prompt's plaintext the tuple is distributed as $H_{i-1}$, and when it encrypts zero as $H_i$. $B_i$ runs in time at most $t + c(g)\,t_{\mathrm{enc}}$, so
$$\bigl|\Pr[D(H_{i-1}) = 1] - \Pr[D(H_i) = 1]\bigr| \le \mathrm{Adv}^{\mathrm{evk}}_g\bigl(t + c(g)\,t_{\mathrm{enc}}\bigr).$$
Summing over $i$ gives the bound, and the computed ciphertexts and the bootstraps add no hybrid, because they are functions of the tuple that $D$ may compute itself.

\paragraph{What the theorem does not cover.} The view contains no decryption, so IND-CPA in the presence of the keys is the notion the theorem needs, and the decryption-aware notion for approximate schemes \cite{limicciancio} does not enter. The theorem covers one run. It extends to several runs under one key set when every continuation is a function of the client's own inputs. The same hybrid then runs over all prompt ciphertexts with the keys of the union of the rotation sets. The bound sums the $c(g_j)$ times the advantage of that larger instance of the assumption. A continuation, a retry or a published token that depends on a decrypted value is a function of the decryption error. The reduction cannot compute that error, so the theorem does not cover such a use. Rounding to a token gives no protection, because the encrypted argmax can differ from the plaintext argmax at near ties. Noise flooding at the output is the known repair \cite{lidpflood}. It needs a certified bound on the noise of the output ciphertext under our parameters, which we do not measure. The threat model therefore assumes that the client acts on no decrypted value where the server observes it, retries included. The wall clock of a run is not in the view, and a public schedule fixes the count of every operation and not the time each one takes. The theorem therefore covers neither timing nor memory access, both of which the threat model places out of scope. It does not cover integrity, since a server that deviates from $P_g$ may return arbitrary ciphertexts that the client cannot detect. It also does not cover model confidentiality against a client with query access, availability, or session metadata such as when a request arrives.
\section{The sites behind the cost law}
\label{sec:supp-constructions}

The coefficients of the cost law count the bootstrap sites of the decode step, where a site is a point at which an operand enters a stage. This appendix lists the sites behind each coefficient and states the three rules the implementation follows at the input of every homomorphic inverse. It then gives the two implementation checks that the experimental setup mentions.

\subsection{The site ledger}
\label{sec:supp-ledger}

The law has three coefficients, $\RhoAtt$, $\RhoLay$ and $\RhoFfn$, and a fourth quantity $\RhoNorm$ that enters through $\RhoLay$. $\RhoAtt$ counts the bootstraps of one pair of a position offset and a state ciphertext on the attention path of one layer. The pair has one bootstrap at the entry of the exponential, one at the normalization of stage 0, one at each of the five square-and-normalize rounds of the softmax, and one on the attention-value path before the product. The normalization of stage 0 is a site that this parameter set adds, because the exponential consumes nine levels and the causal mask one more. The operand therefore reaches the threshold under a longer modulus chain and falls below it under this one. A softmax approximation with another number of rounds gives another $\RhoAtt$, and the coefficient then follows the layer count without further change.

$\RhoLay$ counts the bootstraps that a layer places once per state ciphertext. These are two normalizations, three entry bootstraps for the query, key and value projections, the entries of the output projection and of the first feed-forward chunk, six for the denominator inverses of the softmax at stage 0 and at each of the five rounds, and two for the linear renormalization, one guard and one inverse entry. $\RhoNorm$ counts the bootstraps of one normalization, one at its entry, three inside the Newton iteration for the inverse standard deviation and one at its exit. The exit is a site this parameter set adds, and it occurs at both normalizations of a layer and once more in the tail. $\RhoFfn$ counts the bootstraps of one feed-forward chunk, one before the GELU, one after it and eight inside the Chebyshev evaluation of the GELU polynomial. The lowest arrival level of that evaluation falls below the threshold on this parameter set, so the chunk term is linear in $K$.

One site does not occur uniformly, and the $-(C-W)$ term of the law accounts for it. A state ciphertext not written during the run retains the level at which it was encrypted, so the entry bootstrap of the first normalization is skipped on it. That entry occurs at every state ciphertext in the layers above layer 0 and only at the written ones in layer 0, which gives $(L-1)C + W$ bootstraps against $LC$. The constant term counts what a step places once, the final normalization and the entry of the projection after the last layer. $S$ counts the selection stage under CutMax's argmax.

\subsection{The rules at a homomorphic inverse}

The implementation guards against distribution shift at the input of a homomorphic inverse with three rules. The first rule puts a floor or a validity guard on every homomorphic inverse over all the slots the computation can reach, including the slots beyond those the designer expects. The second rule keeps every intermediate that a bootstrap sees within a fixed range of about $[0.1, 2]$ by constant scaling before and after the operation. The third rule sets the margins from the worst input the circuit can receive. Where that input and the typical benchmark input differ by orders of magnitude, the rule re-parameterizes the circuit. The shift appeared at seven sites, and the rules closed each of them. The sites are the scale at the entry of selection, the floors of the round sums, the pad denominators of the softmax, the fill of the empty blocks in LayerNorm, the validity guard of the softmax, the coupling of the winner's share and the mass in the final renormalization, and the basin of LayerNorm at valid but tiny write-backs. The eighth site, the first inverse of the attention softmax, remains open, and Appendix~\ref{sec:supp-correctness} gives its measurement.

\subsection{Two implementation checks}

Splitting the special modulus in two halves the size of each evaluation key, and we checked that it leaves the error of a bootstrap unchanged. Three arms of twenty isolated calls at ring degree $2^{15}$, with the same usable levels, returned a post-bootstrap error of $2.164240\times10^{-6}$ with two special primes against $2.164191\times10^{-6}$ with one. The error is the same to six digits, and the key set is 240.0 MiB against 465.0 MiB. The number of key-switching digits is the chain length divided by the number of special primes. Two divisibility checks reject a chain that the primes do not divide, because such a chain drops primes from the decomposition and produces a wrong result. The second check concerns the transfer of the keys to the workers of the device axis. Each worker holds forty nine keys, 20.10 GiB, and receives them in 4.2 to 4.3 seconds with one worker and in about 4.1 seconds per worker with two and three.
\section{The Llama-3-8B column}
\label{sec:supp-llama}

The Llama-3-8B column of the evaluation is derived, and this appendix gives the derivation. The model has $d = 4096$, $L = 32$ and a vocabulary of $128{,}256$ tokens, with RMSNorm, SwiGLU, rotary position embeddings and grouped-query attention. At the parameter set of Appendix~\ref{sec:supp-card} it fills nine state ciphertexts at a context of 32. The cost law evaluated on this model gives \LlamaPromptBoots{} bootstraps for a prompt step and \LlamaGenBoots{} for a generated step. A level-by-level simulation of the circuit produces the same two counts. The cached form gives \LlamaCachedBoots{} bootstraps for a cached generated step, which is the generated count divided by \LlamaCacheFactor. The bootstrap duration measured on this parameter set is \RefreshPriceMs{} ms. One device therefore needs \LlamaPromptHours{} hours of bootstraps for a prompt step, \LlamaGenHours{} hours for a generated step and \LlamaCachedHours{} hours for a cached generated step. Because no Llama-3-8B step was measured, each figure is a derived count times a measured duration, and the column has no wall clock of its own.
\section{The device axis in detail}
\label{sec:supp-devices}

The evaluation reports the device axis as a result and as a lower bound. This appendix gives the two runs behind it and the barriers on which the axis acts. In the first run the summed wall clock of the barriers is 42251.7, 40121.6 and 38333.0 ms from two to four devices. In the second it is \DevRepBarrierWallTwo, \DevRepBarrierWallThree{} and \DevRepBarrierWallFour{} ms. Two and four devices replicate within \DevReplicateTwoPct\% and \DevReplicateFourPct\%, whereas three devices do not, because the repetition is \DevThreeExcessPct\% above the first run and above two devices. The two runs therefore do not determine the wall-clock order of two and three devices. In the repetition the width-two barriers account for the excess at three devices. Their mean is \DevRepWidthTwoMeanThree{} ms against \DevRepWidthTwoMeanTwo{} at two devices and \DevRepWidthTwoMeanFour{} at four, and the excess grows as the step proceeds. The first \DevWidthTwoEarlyCount{} of these barriers take \DevWidthTwoEarlyLowMs{} to \DevWidthTwoEarlyHighMs{} ms. From the fourth layer on, the last \DevWidthTwoLateCount{} take \DevWidthTwoLateLowMs{} to \DevWidthTwoLateHighMs{} ms, against \DevGridWidthTwoLateLowMs{} to \DevGridWidthTwoLateHighMs{} ms for the same barriers in the first run. In the repetition the generated step at three devices exceeds the step at two devices by \DevRepThreeStepExcessS{} seconds. Of that excess, \DevRepThreeOutsideBarriersS{} seconds fall outside the barriers, where the master device runs alone. The prompt step has no barrier and takes \DevPromptStepThreeS{} seconds at three devices in both runs, against \DevPromptStepLowS{} to \DevPromptStepHighS{} seconds elsewhere. We do not identify a cause for that difference.

The barrier model of the evaluation predicts ratios of 1.886, 1.062 and 1.000 for the summed barrier wall clock from two to four devices. The measured ratios are 1.1022, 1.0467 and 1.0000 in the first run and \DevRepRatioTwo, \DevRepRatioThree{} and 1.0000 in the second. The model therefore overstates the gain of the second device in both runs. In the first run a token takes 255.6, 236.7, 239.2 and 233.8 seconds from one to four devices. In the second it takes \DevRepSecPerTokenOne, \DevRepSecPerTokenTwo, \DevRepSecPerTokenThree{} and \DevRepSecPerTokenFour{} seconds, and the share of the work that stays on the master device is 8.13\% measured against 7.9\% predicted. We judge the axis by the summed barrier wall clock because the axis changes that term. The wall clock of the whole step separates two, three and four devices by at most \DevStepWallSpreadPct\% across the two runs, against a within-process bound of 1.7\%. The axis removes a fixed set of bootstraps from the critical path whatever their share of the step.
\section{The context boundary and the cache counters}
\label{sec:supp-correctness}

\subsection{The context boundary}

The loop returns the plaintext model's tokens at every prompt length the evaluation ran on nanoGPT except one, and the failure at that length occurs at a single operation. We localized it on the development parameter set, at ring degree $2^{15}$, where the same failure appeared at a prompt of 49 tokens. We decrypted every intermediate value of the diverging step and of a healthy step under the secret key that the measurement program holds. On the paper's parameter set the sweep of the evaluation places the boundary at a prompt of 48 tokens, and we did not repeat the localization there.

The failure occurs at the inverse of the first round of the attention softmax. The inputs of the two steps to that inverse agree to 6\% while their outputs differ by a factor of 56. The magnitude of the operand is therefore not the cause. What separates the two steps is the bootstrap before the inverse. Each probed block contains 128 pad slots that the validity fill sets to 1.0, so every inversion also inverts a known input. Those slots give the additive error of that bootstrap, $-1.7\times10^{-4}$ at the healthy step and $-5.554\times10^{-3}$ at the diverging one. The error is coherent across the slots of a block. The magnitude is specific to the individual bootstrap, because the two readings at the same operation differ by a factor of 32.

The operand of that inverse is confined to a range with a bound at each end. The Goldschmidt iteration $D \mapsto D(2-D)$ converges only below $u = 2$ and diverges past it. The error $\beta$ of the bootstrap must stay a small fraction $f$ of the operand, which requires $u \ge \beta/f$. Together the two bounds give
\begin{equation}\label{eq:window}
u \;\in\; \left[\beta/f,\; 1.4\right], \qquad
\frac{u_{\max}}{u_{\min}} \;=\; \frac{1.4\,f}{\beta} \;=\; 12.6
\end{equation}
at the measured $\beta = 5.55\times10^{-3}$ and $f = 0.05$. Under the deployed scaling the operand of layer 0 decreases as $1/n$ and reaches $\beta$ near $n = 45$ on that parameter set, against the collapse observed at 49. A single scaling constant is admissible only if it maps the denominator of every round in every layer into the range. The spread of those denominators grows with the context, at least as $0.25\,n$, so it exceeds the width of the range near $n = 51$. Raising the scale to restore the margin of layer 0 sends the peaked rows of layer 1 past $u = 2$, so the two ends of the range fail together. A floor that masks the error must be comparable to the operand. A floor of that size perturbs the round by an order-one factor that the four later squarings of the softmax raise to the sixteenth power.

\subsection{How far a generation stays exact}

On the prompt that generated the most tokens, the loop returned the plaintext model's token 24 times out of 24 and reached a context of 55. On a second prompt it returned 16 of 16 and reached a context of 47. On a third it returned 5 of 8, with the first loss at the fifth token. The first of these prompts was the joint best of a screen of twenty prompts, and the other two were not selected. A repetition of the 16-token run under a second secret key returned the same 16 tokens, with 71006 bootstraps in both runs and wall clocks 0.30\% apart.

\subsection{The cache counters}
\label{sec:supp-cache-counters}

We read two counters on every run to check the cache. At both measured models the bootstrap count of the cache was zero on every cached step. Every appended key or value arrived five levels above the level at which its entry is held. A separate run appended to one entry 384 times, and its bootstrap count stayed at zero against five bootstraps in the comparison run. The drift after 64 appends was four orders of magnitude below its budget. The error of the appended terms grows with age at an exponent of \CacheAgingExponent, where independent terms would give 0.5, so the terms are coherent. An extrapolation to older entries has to use the measured exponent.
\section{The projections}
\label{sec:supp-projections}

The last subsection of the evaluation projects the cost of models we did not run as a bootstrap count from a closed form times the bootstrap duration we measure. This appendix gives the form, the check it passes before we project any model, the memory model, the layer split, the stages of a step and the assumptions.

\subsection{The cached form}

The cost law is
\begin{equation}\label{eq:supp-proj-law}
\CostLawGeneral ,
\end{equation}
in the bootstrap counts per attention site, per layer and per feed-forward chunk, the layer count $L$, the chunk count $K$, the state ciphertext count $C$, the written-back count $W$ and the tail $\tau$ of the step. Caching the keys and values changes two terms of it,
\begin{equation}\label{eq:supp-proj-cached}
B_{\mathrm{cached}} = \RhoAtt L C + (\RhoLay + \RhoFfn K) L + \tau_{0} .
\end{equation}
The attention term loses its factor of $n$, because a cached step attends over the stored keys and computes the key and value of one position only. The per-layer term loses its factor of $C$, because a cached step runs the layers over the new position alone. The tail $\tau_{0}$ is the tail of the step without the write-back bootstrap, which a cached step does not need.

\subsection{The check against the measured models}

We project a model only after Equation~\ref{eq:supp-proj-cached} has returned both measured cached counts, \CachedStepBoots{} at nanoGPT and \GptTwoCachedStepBoots{} at GPT-2 small. The two models differ in width, in layer count, in vocabulary and in state ciphertext count, so the check exercises every factor of the form. The count check of the evaluation covers the uncached law over the whole context sweep.

\subsection{Memory on one device}

The memory footprint of a run on one device consists of five terms. The evaluation keys are the largest, at \ProjKeySetGiB{} GiB for the key count we measure. The runtime holds the context, the tables, the key pair and one bootstrap in flight. The weights take almost no memory, because the forward pass encodes each diagonal when it uses it. The working set consists of the operands a step keeps live, and the cache takes \ProjParkPairMiB{} MiB for the key and the value of one position in one layer. A layer split divides only the cache, because each device stores the entries of its own layers. The key term is a lower bound, since it is the count measured at nanoGPT and a wider model needs at least as many rotation keys. Every device count the model returns is therefore a lower bound. Against the two models whose whole-loop footprint we measure, the memory model is within about \ProjMemErrPct{} per cent, low at one and high at the other.

\subsection{The layer split}

A layer split changes no bootstrap site and therefore keeps the bootstrap count, and sequential layers keep the per-token critical path. At each device boundary the split adds one transfer of the state ciphertexts. That transfer takes \ProjTransferMs{} ms at the \ProjTransferRateGiBs{} GiB per second measured for the key transfer, against a bootstrap of \RefreshPriceMs{} ms. Throughput then scales by up to the device count over independent requests in flight. The device axis that the evaluation measures is the other placement, and its lower bound remains as measured.

\subsection{The stages of a step}

\begin{figure}[t]
\centering
\includegraphics{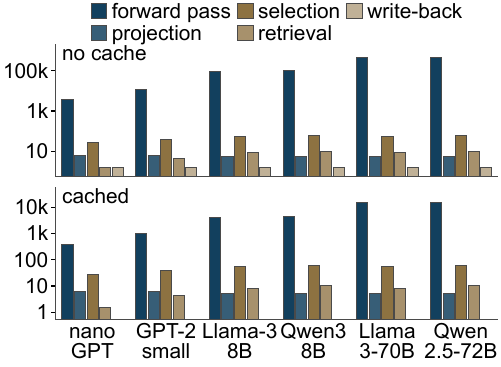}
\caption{Bootstraps of one step, split into five stages at a context of 32, without the cache (top) and with it (bottom). The figure omits Mistral-7B, whose step agrees with Llama-3-8B's to three hundredths of a percent.}
\Description{Two panels of grouped bars on a logarithmic bootstrap axis, six models across and five stages in each group, coloured on one blue to tan scale, without the cache above and with it below.}
\label{fig:projstages}
\end{figure}

Figure~\ref{fig:projstages} divides a step into its five stages for each projected model, without the cache and with it. The forward pass accounts for most of the step at every model, and a cached step has no write-back.

\subsection{The assumptions}

Each projection rests on the following assumptions, and each assumption fixes the direction of its error. The projected counts include the selection term of the level simulator and no further correction. At a model whose selector runs the multi-ciphertext path they are therefore about three bootstraps low out of thousands. The coefficients are read at a context of 32 and applied at every context. The closed form agrees with the simulator at 64 and at 128, beyond which no run tests it. The key count at every model is the count measured at nanoGPT, a lower bound, so every device count is a lower bound. The fixed cost of the runtime is read once from the budget of the parameter set and is not measured per model. The cached working set is counted at nine operands per state ciphertext, which is conservative, because the query side of a cached step is one position wide. The layers divide evenly across the devices, only the state ciphertexts cross a device boundary and requests in flight do not contend, so the device count is an upper bound. The bootstrap duration is measured on this parameter set and does not depend on the model. Because a second reading of it differs from the first, we defend no second significant figure of a projected time. The projections cover dense models with full attention.
\section{The results at row level}
\label{sec:supp-tables}

Table~\ref{tab:counts} lists the predicted and measured bootstrap counts of the evaluation at row level. The comparison record below gives the readings behind the comparison against NEXUS and against the client-interactive alternative.

\begin{table}[t]
\centering
\footnotesize
\setlength{\tabcolsep}{3pt}
\caption{Predicted and measured bootstrap counts on nanoGPT under CutMax's argmax, and at the two models that played no part in deriving the formula. The upper block lists eight of the twelve contexts of the sweep. These are the two crossings marked $\dagger$ with the context before each, where the state ciphertext count rises by one, the anchor pair at contexts 32 and 33, read at least 60 times on at least six nodes, and the two ends of the sweep. The generated step of each row is at the context one greater than the prompt step's. No measured count differed from its prediction at any of the 24 combinations of context and step type, over 136 readings. The Llama-3-8B row is derived, since no Llama-3-8B step was measured.}
\label{tab:counts}
\begin{tabular}{@{}lcrrrr@{}}
\toprule
& & \multicolumn{2}{c}{Prompt step} & \multicolumn{2}{c@{}}{Generated step} \\
\cmidrule(lr){3-4} \cmidrule(l){5-6}
Context & $C$ & Predicted & Measured & Predicted & Measured \\
\midrule
\multicolumn{6}{@{}l}{\textit{The formula across the context, nanoGPT}} \\
16 & 1 & 1107 & 1107 & 1157 & 1157 \\
30 & 1 & 1779 & 1779 & 1829 & 1829 \\
31$^\dagger$ & 2 & 3620 & 3620 & 3717 & 3717 \\
32 & 2 & 3716 & 3716 & 3814 & 3814 \\
33 & 2 & 3812 & 3812 & 3910 & 3910 \\
62 & 2 & 6596 & 6596 & 6694 & 6694 \\
63$^\dagger$ & 3 & 10021 & 10021 & 10166 & 10166 \\
64 & 3 & 10165 & 10165 & 10311 & 10311 \\
\midrule
\multicolumn{6}{@{}l}{\textit{Models outside the derivation, at $n=32$}} \\
GPT-2, 65 tokens & 3 & 11443 & 11443 & 11733 & 11733 \\
GPT-2, full vocab. & 3 & 11458 & 11461 & 11748 & 11751 \\
Llama-3-8B & 9 & 90{,}779 & & 93{,}085 & \\
\bottomrule
\end{tabular}
\end{table}

\subsection{The packed count across the context}
\label{sec:supp-packed-flat}

At the widest group the measured packed count is 387 at a context of 32 and 740 at contexts of 40, 48 and 64. At half that group width it is 788 at contexts of 40 and 48. Each block of counts is constant over the contexts it covers, and the difference between the blocks is the state ciphertext count.

\subsection{The comparison record}
\label{sec:supp-comparison-record}

The comparison uses NEXUS's own parameters, so its numbers belong in no cost column of this paper. Our GELU takes 70.35 ms, 1.020 of NEXUS's reading, and our LayerNorm 43.74 ms, 0.972 of it, each over 30 calls with a deviation under 0.5\%. Our matrix multiplication takes 0.2993 s per column on an H100 against the 1.31 s per column NEXUS reports for an A100. Our softmax microbenchmark takes NEXUS's own input, and its decoded output equals NEXUS's up to one common scalar of \SoftmaxScalar. After that scalar is divided out, its error is \SoftmaxPostScalarFloorRatio{} of NEXUS's own error against its calibration, and the origin of the scalar remains unidentified. An earlier measurement of the matrix multiplication was taken before the output check existed, so we withdraw it, and we compute no difference against the figure we withdrew. The multi-device variant of the matrix multiplication fails that check on the current build, at \MatmulMultiDevErr{} against a tolerance of \MatmulTol. The single-device path passes at \MatmulSingleDevErr, so the paper reports the single-device figure.

The interactive loop is the alternative in which the server sends the encrypted logits to the client, who decrypts them, selects the token and returns its embedding encrypted. We derive its cost on our own implementation with every assumption in the alternative's favour, and every input to the derivation is a measurement of our own. The alternative takes 1062.0 seconds per generated token against our measured \GptTwoSecPerToken, a saving of 2.216 percent. It requires 4.72 hours of client presence at sixteen tokens against two transfers for our loop.

\end{document}